\documentclass[useAMS,usenatbib]{mn2e}
\usepackage{graphicx}
\usepackage{color}
\usepackage{epsfig}
\usepackage{amsmath,amssymb}
\usepackage{natbib}
\usepackage{comment}
\usepackage{multirow}
\usepackage{lscape}
\usepackage{threeparttable}
\usepackage{bm}
\usepackage{ulem}

\def\simge{
    \mathrel{\rlap{\raise 0.511ex
        \hbox{$>$}}{\lower 0.511ex \hbox{$\sim$}}}}
\def\simle{
    \mathrel{\rlap{\raise 0.511ex
        \hbox{$<$}}{\lower 0.511ex \hbox{$\sim$}}}}

\newcommand{\figref}[1]{Fig.~\ref{#1}}
\newcommand{\tabref}[1]{Table~\ref{#1}}
\newcommand{\secref}[1]{Section~\ref{#1}}

\newcommand{\cm}{{\rm cm}}
\newcommand{\pc}{{\rm pc}}
\newcommand{\msun}{{\rm M}_{\odot}}
\newcommand{\mbh}{M_{\rm BH}}
\def\mbhnum#1{M_{\rm BH,#1}}
\newcommand{\mstar}{m_*}
\def\mstarnum#1{m_{\rm *,#1}}

\newcommand{\rsun}{R_{\odot}}
\newcommand{\rstar}{r_*}
\def\rstarnum#1{r_{\rm *,#1}}

\newcommand{\mclb}{M_{\rm cl,b}}
\newcommand{\mclbz}{M_{\rm cl,b0}}

\newcommand{\yr}{{\rm yr} }

\newcommand{\Myr}{{\rm Myr} }
\newcommand{\Gyr}{{\rm Gyr} }
\newcommand{\Mpc}{{\rm Mpc} }
\newcommand{\Gpc}{{\rm Gpc} }

\newcommand{\kms}{{\rm km\,s^{-1}}}
\newcommand{\mc}[1]{\multicolumn{1}{c}{#1}}

\usepackage{xparse,soul}
\makeatletter
\NewDocumentCommand{\nred}{O{red}O{red}+m}
    {%
        \begingroup
        \setulcolor{#1}%
        \setul{-.5ex}{.4pt}%
        \def\SOUL@uleverysyllable{%
            \rlap{%
                \color{#2}\the\SOUL@syllable
                \SOUL@setkern\SOUL@charkern}%
            \SOUL@ulunderline{%
                \phantom{\the\SOUL@syllable}}%
        }%
        \ul{#3}%
        \endgroup
    }
    
\NewDocumentCommand{\nblue}{O{blue}O{blue}+m}
    {%
        \begingroup
        \setulcolor{#1}%
        \setul{-.5ex}{.4pt}%
        \def\SOUL@uleverysyllable{%
            \rlap{%
                \color{#2}\the\SOUL@syllable
                \SOUL@setkern\SOUL@charkern}%
            \SOUL@ulunderline{%
                \phantom{\the\SOUL@syllable}}%
        }%
        \ul{#3}%
        \endgroup
    }
\makeatother

\title[Growth of IMBHs in the first star clusters]
{Growth of intermediate mass black holes by tidal disruption events in the first star clusters}

\author[Y. Sakurai, N. Yoshida and M. S. Fujii]
{Yuya Sakurai$^{1}$\thanks{yuya.sakurai@physics.gatech.edu}, 
Naoki Yoshida$^{2,3,4}$, Michiko S. Fujii$^{5}$
\\ \\
$^{1}$School of Physics, Georgia Institute of Technology, Atlanta, GA 30332, US\\
$^{2}$Department of Physics, School of Science, The University of Tokyo, 7-3-1 Hongo, Bunkyo, Tokyo 113-0033, Japan\\
$^{3}$Kavli Institute for the Physics and Mathematics of the Universe (WPI), UT Institute for Advanced Study, The University of Tokyo, \\
~Kashiwa, Chiba 277-8583, Japan \\
$^{4}$Research Center for the Early Universe (RESCEU), School of Science, The University of Tokyo, 7-3-1 Hongo, Bunkyo, \\
~Tokyo 113-0033, Japan\\
$^{5}$Department of Astronomy, School of Science, The University of Tokyo, 7-3-1 Hongo, Bunkyo, Tokyo 113-0033, Japan
}
\begin{document}

\date{Draft version \today}

\maketitle

\label{firstpage}

\voffset=-0.4in

\begin{abstract}
  We study the stellar dynamics of the first star clusters 
  after intermediate-mass black holes (IMBHs) are formed via runaway stellar 
  collisions. 
  We use the outputs of cosmological simulations of \citet{Sakurai2017}
  to follow the star cluster evolution in a live dark matter (DM) halo.
  Mass segregation within a cluster promotes massive stars to be 
  captured by the central IMBH occasionally, causing tidal disruption events (TDEs). 
  We find that the TDE rate scales with the IMBH mass as 
  $\dot{N}_{\rm TDE}\sim0.3\,\Myr^{-1}(M_{\rm IMBH}/1000\,\msun)^2$. 
  The DM component affects the star cluster evolution
  by stripping stars from the outer part. 
  When the DM density within the cluster increases, the velocity dispersion
  of the stars increases, and then the TDE rate decreases.
  By the TDEs, the central IMBHs grow to as massive as $700-2500\,\msun$ in 15 million years.
  The IMBHs are possible seeds for the formation
  of supermassive BHs observed at $z\gtrsim 6-7$, 
  if a large amount of gas is supplied through galaxy mergers and/or large-scale gas accretion,
  or they might remain as IMBHs from the early epochs to the present-day Universe.
\end{abstract}

\begin{keywords}
intermediate mass black holes -- galaxies: star clusters -- stellar dynamics
\end{keywords}

\section{Introduction}
\label{sec:introduction}
A number of quasars
have been discovered at redshift $z\gtrsim6$
\citep[e.g.,][]{Mortlock2011,Wu2015,Banados2018}. 
They are thought to be powered by accreting SMBHs of mass $\gtrsim10^9\,\msun$,
which need to be rapidly assembled within $1\,\Gyr$ after the Big Bang.
The origin of the SMBHs remains largely unknown, but
there are a few promising models of BH formation.
For example, a very massive Population III star may leave a remnant
BH with mass $\sim100\,\msun$ \citep{Madau2001,Haiman2001,Schneider2002}.
Another model considers an even more massive seed BH as a remnant of
a supermassive star (SMS) with mass $\gtrsim10^5\,\msun$ \citep{Loeb1994,Oh2002,Bromm2003}.
The growth of such seed BHs is also under debate.
Gas accretion onto an early BH is likely to be suppressed by radiation feedback effects \citep{Alvarez2009, Jeon2012}, whereas it is also possible that 
super-Eddington accretion is achieved under large gas mass accretion
(\citealt{Pacucci2015}; \citealt{Inayoshi2016,Sakurai2016}).

It has been suggested that an intermediate-mass BH (IMBH) 
with mass $\sim1000\,\msun$ can be formed
in a dense star cluster via runaway stellar collisions
(\citealt{Ebisuzaki2001}; \citealt{PortegiesZwart2002, PortegiesZwart2004,Gultekin2004}; \citealt{Vanbeveren2009, Devecchi2010}; \citealt{Giersz2015, Stone2017, Reinoso2018, Kovetz2018}).
Dense star clusters can be formed in the early universe by
fragmentation of low-metallicity gas clouds \citet{Omukai2008}. 
\citet{Katz2015} performed cosmological simulations and N-body simulations
to follow the evolution of early star clusters. 
They show that the runaway collisions occur within the star clusters
of $\sim10^4\,\msun$ in low-mass 
($\sim10^6\,\msun$) dark matter haloes.
They also show that very massive stars of several hundred solar mass
are formed at the cluster center, which leave IMBHs by gravitational collapse.
\citet{Sakurai2017} perform cosmological simulations to locate host dark matter halos
of the first star clusters, and study the evolution of a number of star clusters
in atomic cooling halos at $z\sim12-20$.
In all the eight star clusters in their simulations, very massive stars with mass 
$\sim400-1900\,\msun$ are formed via stellar collisions.

It is worth studying the later evolution of the first star clusters
because tidal disruption events (TDEs) are expected to occur when
stars approach the central IMBH.
Intriguingly, the TDEs can be bright enough to be observed by SwiftBAT and
eROSITA even 
if a star of $\lesssim100\,\msun$ is disrupted by an IMBH of $\sim10^5\,\msun$
\citep{Kashiyama2016}; the peak luminosity of the TDEs will be even higher for smaller BHs.

In the present paper, we study the dynamical evolution of the first star clusters
after the formation of IMBHs. 
We run $N$-body simulations to follow the cluster evolution for $\sim15\,\Myr$ 
which is comparable to the typical relaxation time of a very dense cluster.
The clusters are expected to evolve substantially over a relaxation time.
We show that the central IMBHs grow by TDEs and that the TDE rate scales as 
square of the IMBH mass.
The cluster evolution depends on the details of the DM distribution
and relative motions between the cluster and the DM halo;
stars are stripped from the outer part of the cluster.

We organize the rest of the paper as follows. 
In \secref{sec:method}, we describe the initial conditions of our simulations and 
the details of our $N$-body simulations.
In \secref{subsec:IMBH mass growth}, we study mass growth of the IMBHs by TDEs.
In \secref{sec:global_ev}, we study how the DM halo affects
the cluster evolution. 
In Sections \ref{sec:discussion} and \ref{sec:summary}, we discuss and summarize our results.

\section{Numerical Methods}
\label{sec:method}
\subsection{Initial conditions of primordial star clusters from \citet{Sakurai2017}}
\label{subsec:IC}
The initial conditions are generated from the final states of the cluster simulations studied in \citet{Sakurai2017} (hereafter SYFH17). We use the eight models A-H,  
and three realizations for each model.
Hereafter, we refer the initial time $t=0$ when the IMBHs are born after 3 million years elapsed
  since the beginning of our cluster simulations.

We briefly summarize the physical properties of these clusters (see table 1 of 
SYFH17).
The gas clouds that produce star clusters are located at $z \sim12-20$
in atomic-cooling halos. 
The haloes have masses of
$(1.5-4.2)\times10^7\,\msun$.
In the cluster generation procedures at $t=-3\,\Myr$, 
\begin{itemize}
\item SPH particles are replaced with star particles in a probabilistic manner
  based on Equation (1) in SYFH17 which is determined to preserve mass conservation
  and to yield a global star 
  formation efficiency of $\sim5-10\%$, 
\item the stellar mass distributions are determined by Salpeter initial mass function (IMF) with the minimum mass $3\,\msun$ and the maximum mass $100\,\msun$,
\item the DM particles are retained as in the original zoom-in simulations with a DM particle mass $1.87\,\msun$ and
\item the velocities of the star and DM particles are rescaled
  such that the star cluster systems are approximately in virial equilibrium.
\end{itemize}
The resulting star cluster mass ranges around $(5-16)\times 10^4\,\msun$,
with the number of stars being $(6-20)\times10^3$ and the core radius
$\sim0.2-0.7\,\pc$ \citep{Casertano1985}. 
The star cluster systems have total DM masses of $\sim(3-7)\times10^7\,\msun$ and the number of DM particles of $\sim(2-4)\times10^7$.
To follow the dynamical evolution and runaway stellar collisions within the clusters, 
a hybrid $N$-body simulation code BRIDGE \citep{Fujii07} is used.
The evolution is followed for $3\,\Myr$,
which is approximately equal to the lifetime of massive stars.
In most of the simulations of SYFH17, runaway stellar collisions occur
and massive stars with masses 
$\sim400-1900\,\msun$ are formed.
Such very massive stars are expected to collapse gravitationally to IMBHs
with the same masses.

\begin{figure*}
    \centering
    \includegraphics[width=0.7\textwidth]{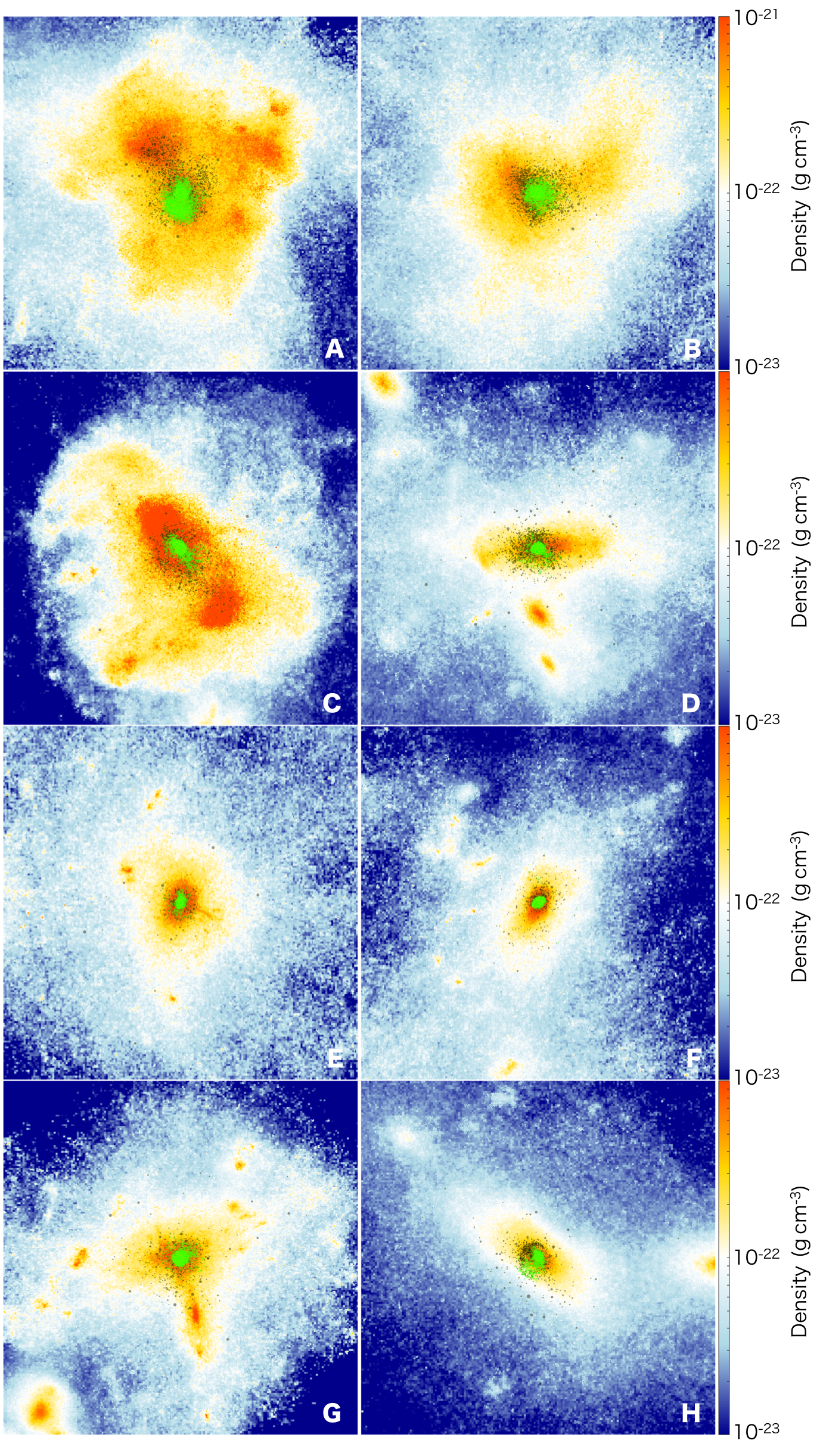}
    \caption{Initial distributions of the stars and DM
      within $400\,\pc$ on a side,
      projected along $z$-axis.
    The green dots represent bound stars whereas the black dots are unbound stars.
    DM density is shown by colors,
    and the color-bar indicates the density.
    }
    \label{fig:posall}
\end{figure*}

In the present study, we replace the central very massive stars with IMBHs, and follow the
dynamical evolution for further 15 million years.
Each model cluster has one IMBH at the center.
In \tabref{tab:SC}, we summarize the initial properties of our star cluster samples.
\figref{fig:posall} shows the initial spatial distributions of the stars
in our eight models.
For the initial conditions, the stars are considered to be `bound' if the stellar energy 
without DM potential is negative, i.e., 
\begin{equation}
\frac{v_{\rm star}^2}{2}+\phi_{\rm star}<0,
\label{eq:bound cond}
\end{equation}
where $v_{\rm star}$ is the velocity of a star
and $\phi_{\rm star}$ is the gravitational potential generated by only the stars.
For comparison, we also generate an initial condition without DM for 
Model A of \citet{Sakurai2017}. 
The properties of the models are  given in \tabref{tab:SC}.

\begin{table*}
  \begin{center}
    \caption{Initial properties of the star cluster models.
      The model data are taken from the snapshots at $t\sim3\,\Myr$
      of the star clusters studied in \citet{Sakurai2017}.
      For each model, all listed quantities are averaged among three realizations.
    }    
      \begin{tabular}{lcccccccc} \hline
      Model & \mc{$M_{\rm cl}$}      &       \mc{$N_{\rm star}$}    &     \mc{$r_{\rm c}$}   & \mc{$r_{\rm hm}$} &    \mc{$\rho_{\rm c}$}  &  \mc{$t_{\rm rh}$} & \mc{$M_{\rm IMBH,i}$} & \mc{$M_{\rm DM}$}  \\
                 & ($10^4\,\msun$)& ($10^3$)& ($\pc$)& ($\pc$)& ($10^7\,\msun/\pc^3$) & ($\Myr$) & ($\msun$) & ($10^7\,\msun$) \\
                 \hline \hline
A & 16.4  & 19.9  & 0.151 & 1.48 &  3.18 &  16.8 & 917 & 4.79\\
B & 13.0  & 15.7  & 0.132 & 0.887 & 3.35  & 15.4 & 409 & 3.78\\
C & 12.1  & 14.7  & 0.137 & 0.972 &  15.2  & 15.4 & 1312 & 6.60\\
D & 11.7  & 14.1  & 0.153 & 1.03 &  7.89  & 14.9 & 971 & 5.67\\
E & 4.76  & 5.76  & 0.142  & 0.883 & 2.28  & 11.7 & 724 & 3.25\\
F & 9.00  & 10.8  & 0.136  & 0.908 & 12.6  & 14.0 &908 & 5.13\\
G & 12.5  & 15.0  & 0.115 & 0.954 &  25.4  & 15.7& 1628 & 4.17\\
H & 7.70  & 9.32  & 0.159  & 1.13 & 5.15  & 12.9& 964 & 5.25\\
\hline \hline
AnoDM & 16.4 & 19.9 & 0.136 & 1.16 & 9.35 & 16.3 & 874& --- \\
\hline
    \end{tabular}
    
    \label{tab:SC}
    \begin{tablenotes}
      \small
    \item Column 2: total mass of the star cluster, Column 3: total number of stars, Column 4: core radii, 
    Column5: half-mass radii, Column 6: core density, Column 7: half-mass relaxation time, 
    Column 8: initial IMBH mass, Column 9: host halo DM mass. 
    The core radii and the core density are computed using the method described in \citet{Casertano1985}. 
    The core radii, half-mass radii, core density and half-mass relaxation time are calculated using bound stellar particles.
    \end{tablenotes}
  \end{center}
\end{table*}

\subsection{Hybrid N-body simulations}
\label{subsec:sim}
We use the parallel hybrid $N$-body code BRIDGE \citep{Fujii07}
to follow the stellar dynamics within the star clusters.
The hybrid code computes the gravitational forces of stars
by direct calculation using the sixth-order Hermite integrator \citep{Nitadori08}
self-consistently with
the forces of DM calculated by the tree method \citep{Barnes1986} in 
which a second-order leapfrog integrator is used.
In the Hermite integration scheme, the individual timesteps are adopted
to improve the computational efficiency.
In order to increase the parallel efficiency, the NINJA scheme is used \citep{Nitadori2006}.
The Phantom-GRAPE library is adopted to accelerate the computation of the forces 
\citep{Tanikawa2013}.
Parameters of the $N$-body simulations are the same as in table 2
of \citet{Sakurai2017} otherwise noted.
Specifically, we use softening lengths for stars/IMBHs $\epsilon_{\rm cl}=0$ and DM $\epsilon_{\rm DM}=0.0313\,\pc$ in all simulations but ones with a supernova model (see \secref{sec:length}). 
In the latter simulations, we suppress interactions between stars and compact objects by increasing $\epsilon_{\rm cl}$ to $0.0078\,\pc$ in order to make computation feasible.
In our code \citep{Fujii2012}, we do not use any regularization method for hard binaries.
We can, nevertheless, avoid the increase of the energy error when following stellar and star-IMBH collisions, since a star-IMBH hard binary prevents the formation of other hard binaries;
even if other hard binaries are formed, they are immediately broken up by the star-IMBH binary. In addition, a star-IMBH system can merge due to the tidal disruption before it becomes too tight to integrate the Hermite scheme.
\color{black}

The conditions for merger of two stars is set by the so-called
sticky sphere approximation \citep[e.g.,][]{Gaburov2010}:
a pair of stars merge when its separation $d$ becomes less than the sum of the two stellar radii
$r_1+r_2$.
The values of the stellar radii are derived from a fitting formula of \citet{Tout1996}
for zero-age-main-sequence stars.
Although the stellar mergers may occur disruptively with non-negligible mass loss, 
we set the merged stellar mass to the sum of the two stellar mass for simplicity.

TDEs by the IMBHs are also implemented.
When the separation between a star and an IMBH is less than the TDE radius
$r_{\rm t}\equiv 1.3r_*(M_{\rm BH}/2\mstar)^{1/3}$ \citep{Kochanek1992},
where $\mstar$ is the stellar mass, the star is assumed to be disrupted.
Note that we did not use the Schwarzschild radius of a BH
for the merger condition because it is much less than the TDE radius. 
For example, for an IMBH with mass of $10^3\,\msun$ and a star with mass
of $10^2\,\msun$ and a radius $\sim10\,\rsun$, the two radii are 
\begin{align}
R_{\rm Sch}&=\frac{2G\mbh}{c^2}=3.0\times10^8\,\mbhnum{3}\,\cm \\
r_{\rm t}&=1.5\times10^{12}\,
\mbhnum{3}^{1/3}
\mstarnum{2}^{-1/3}
\rstarnum{1}\,\cm, 
\end{align}
where $\mbhnum{x}=\mbh/10^x\,\msun$, $\mstarnum{x}=\mstar/10^x\,\msun$ and $\rstarnum{x}=\rstar/10^x\,\rsun$.
We assume that the stellar mass is all added to the IMBH after the merger.
More realistically, a part of the stellar mass can be ejected depending on the stellar orbit 
and thus may not contribute to the IMBH mass increase
\citep[][see \secref{sec:mass increase} for discussion]{Rees1988}.

We follow the star cluster evolution for $15\,\Myr$, which is near one half-mass
relaxation time
(see \tabref{tab:SC})
\begin{equation}
t_{\rm rh}= \! \frac{0.651\,{\rm Gyr}}{\ln(\gamma N)}
\frac{1\,\msun}{\overline{m}_{\rm *}}
\left(\frac{M_{\rm cl}}{10^5\,\msun}\right)^{1/2}
\left(\frac{r_{\rm h}}{1\,{\rm pc}}\right)^{3/2}, \label{eq:t_rh}
\end{equation} 
where $\gamma\sim0.015$ \citep{Giersz1996,Gurkan2004}, $\overline{m}_{\rm *}$ is the mean stellar mass of the Salpeter IMF, $M_{\rm cl}$ is a cluster mass and $r_{\rm h}$ is a half-mass radius.
Although the time $15\,\Myr$ is much longer than the lifetimes of massive stars $\sim2-3\,\Myr$, 
we do not consider gravitational collapses or supernovae in our main simulations.
We will discuss the impact of the collapse (compact remnant formation) and
  supernovae in \secref{sec:length}.

\section{Results}
\label{sec:results}
\subsection{IMBH mass growth by TDEs}
\label{subsec:IMBH mass growth}
In this section, we describe the details of IMBH mass growth by TDEs in the first clusters. 
We confirm that the IMBHs grow only moderately, to have final IMBH masses $M_{\rm IMBH,f}$
determined by the properties of the host haloes which are
about $1-2\%$ of the cluster mass. 
We also find that the TDE properties are consistent with
those found in literatures \citep[e.g.,][]{Baumgardt2004}.

\begin{figure}
    \centering
    \includegraphics[width=0.45\textwidth]{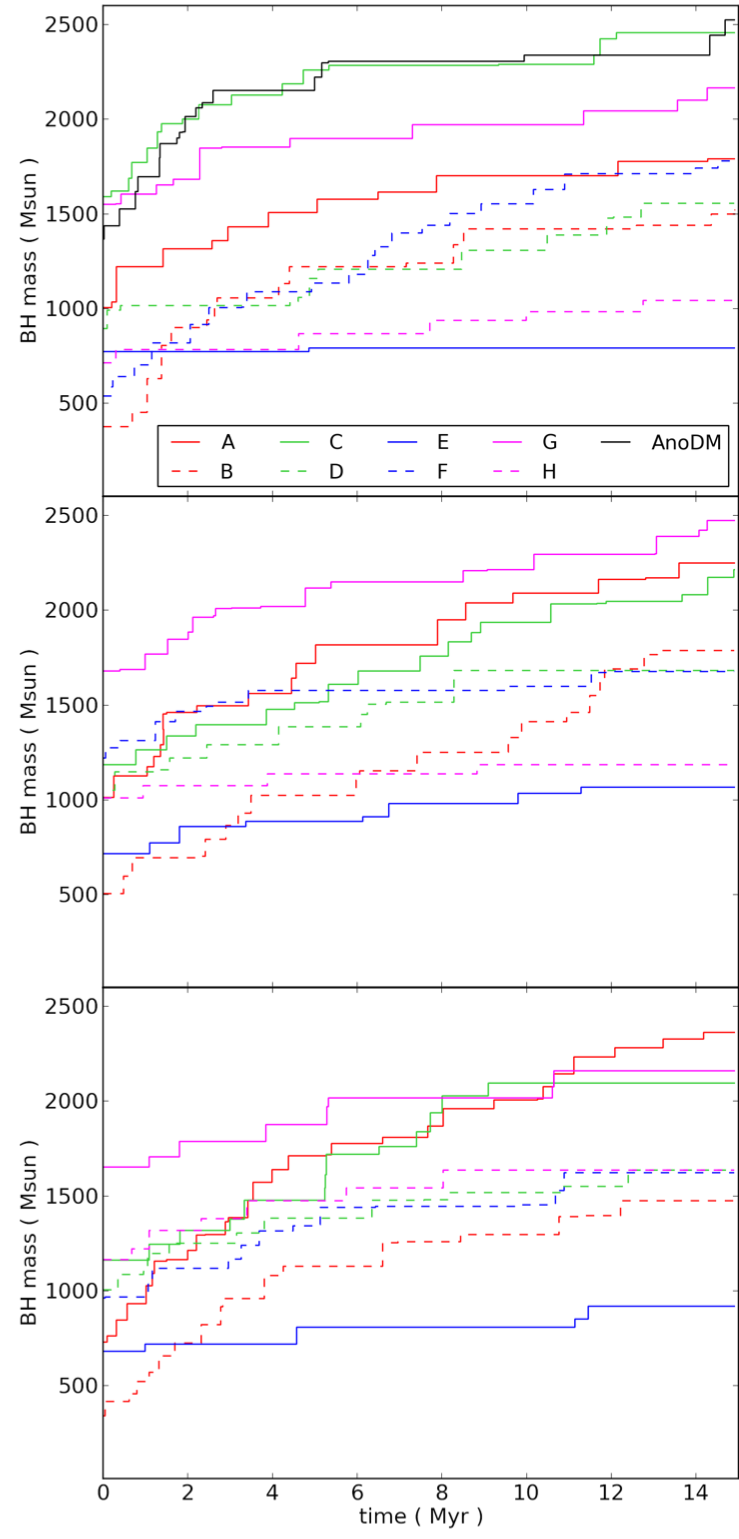}
    \caption{We plot the mass growth of the IMBHs by TDEs in the clusters for all our models. 
    Three panels correspond to three different realizations.
    The initial time $t=0$ is the formation time of the IMBHs.
    }
    \label{fig:t-MBH}
\end{figure}

\begin{figure}
    \centering
    \includegraphics[width=0.5\textwidth]{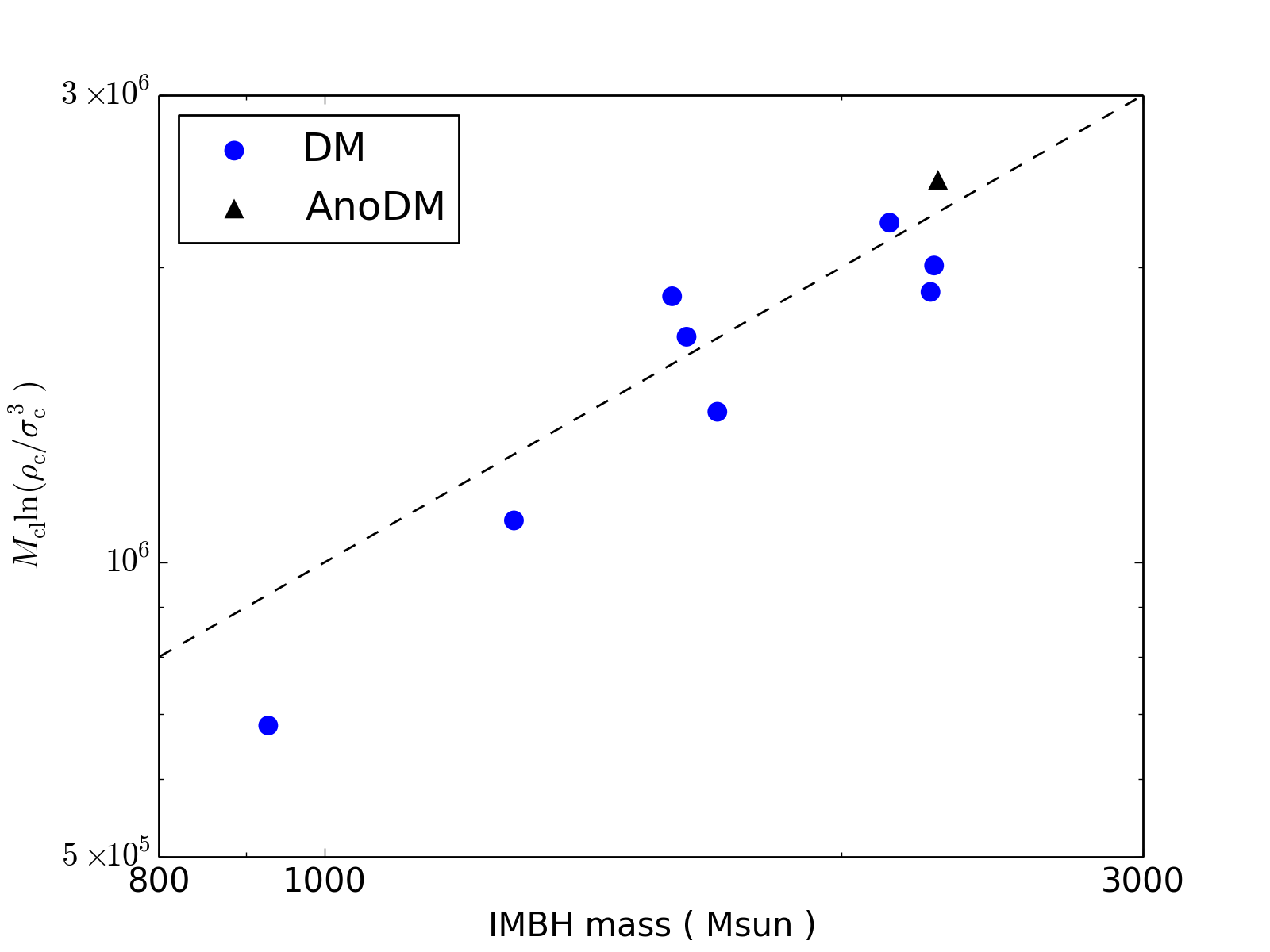}
    \caption{We compare the IMBH mass and a quantity $M_{\rm cl}\ln(\rho_{\rm c}/\sigma_{\rm c}^3)$ from our simulations (points), where $M_{\rm cl}$, $\rho_{\rm c}$ and $\sigma_{\rm c}$ are
      in units of $\msun$, $\msun\,\pc^{-3}$ and $\kms$ respectively. 
    The dashed line is a linear relation of $1000M_{\rm IMBH,f}$ 
    with $M_{\rm IMBH,f}$ by $\msun$ (see equation 5 of \citealt{Sakurai2017}).
    }
    \label{fig:mbh-Mcl}
\end{figure}

\begin{figure}
    \centering
    \includegraphics[width=0.5\textwidth]{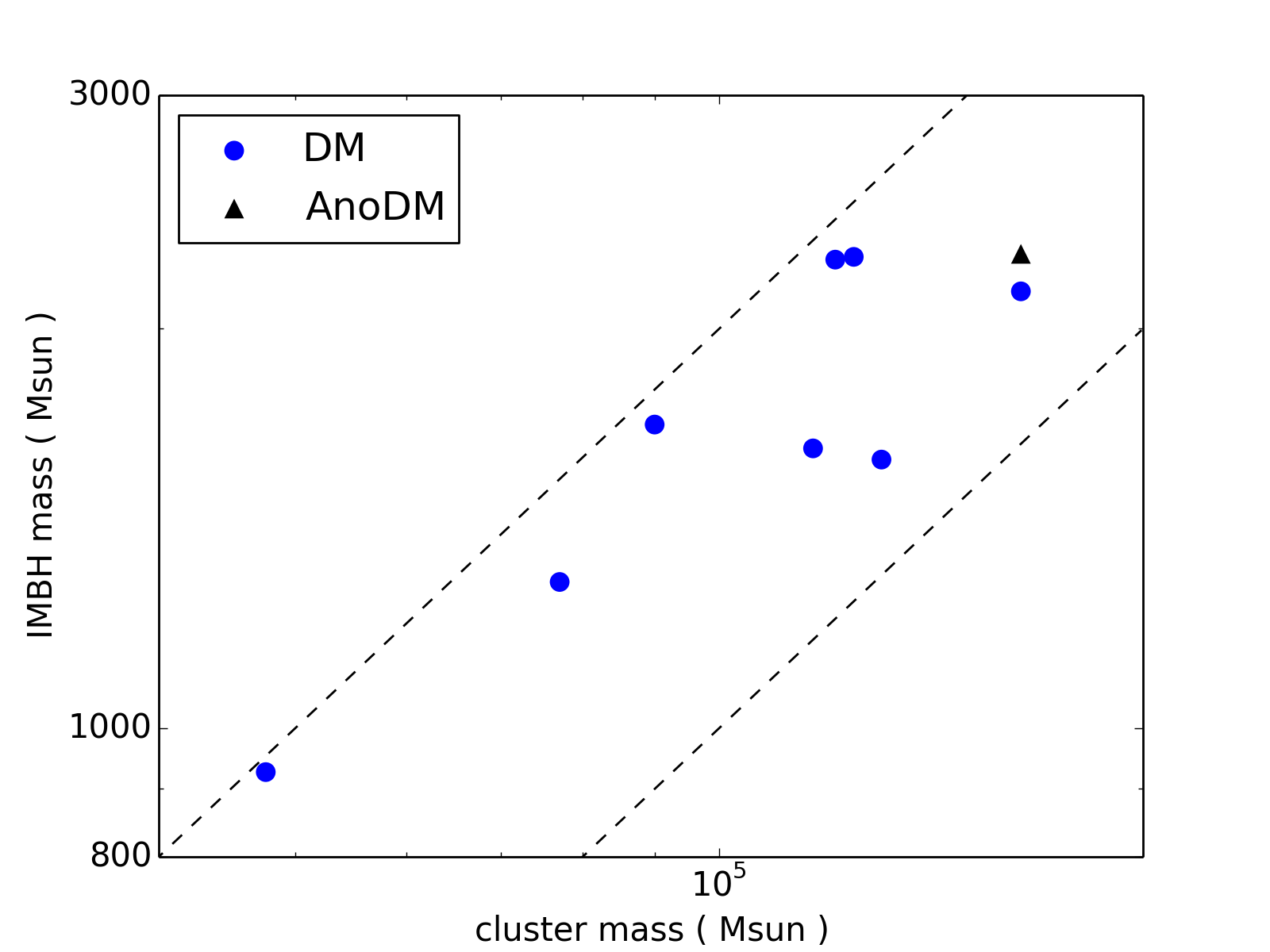}
    \caption{We compare the cluster mass and the IMBH mass from our simulations (points). 
    The dashed lines are linear relations of $M_{\rm IMBH,f}=0.01M_{\rm cl}$ (lower line) and $0.02M_{\rm cl}$ (upper line).
    }
    \label{fig:Mcl-mbh}
\end{figure}

In \figref{fig:t-MBH}, we show the mass evolution of the IMBHs which grow by TDEs.
Three panels correspond to three different realizations.
In \tabref{tab:TDEnum}, we show the final mass of the IMBHs at the end of the simulations.
The IMBHs in the star clusters grow to become as massive as $700-2500\,\msun$ by TDEs.
The diversity of the final IMBH masses can be attributed to the diversity of the properties
of the host clusters, or equivalently, the host haloes (see SFYH17).
\figref{fig:mbh-Mcl} shows that the IMBH mass and a quantity $M_{\rm cl}\ln(\rho_{\rm c}/\sigma_{\rm c}^3)$ \citep{Sakurai2017} linearly correlates with each other, 
where $\rho_{\rm c}$ and $\sigma_{\rm c}$ are the core density and
the core velocity dispersion, respectively. 
The quantity $M_{\rm cl}\ln(\rho_{\rm c}/\sigma_{\rm c}^3)$ is derived by
integrating the mass growth 
rate of the IMBH which undergoes runaway collisions \citep{PortegiesZwart2002}.
In \figref{fig:Mcl-mbh}, we also show the cluster mass-IMBH mass relation
derived from our simulations (blue circles). 
The dashed lines are the relations $M_{\rm IMBH,f}=0.02M_{\rm cl}$ (upper line)
and $0.01M_{\rm cl}$ (lower line).
The final masses are roughly proportional to the cluster mass as $M_{\rm IMBH,fin}\propto M_{\rm cl}$.
We will compare the result with other studies in \secref{sec:evaporation} to explore the impact of DM on the relation and possible scenario of the observed IMBH formation.

\begin{table*}
  \begin{center}
    \caption{Summary of the results of the simulations for $t<15\,\Myr$. 
    }    
      \begin{tabular}{cccccccc} \hline
      Model & $M_{\rm IMBH,f}$ &  $N_{\rm TDE}$ & $\dot{N}_{\rm TDE}$ & $\langle \delta m\rangle$ & $\delta m_{\rm min}$ & $\delta m_{\rm max}$ \\
                 & ($\msun$) &            & ($\Myr^{-1}$) & ($\msun$) & ($\msun$) & ($\msun$)  \\
                 \hline \hline
A &  2134 &    20.0 &  1.33 &  62.2  & 3.21 &  145 \\
B &  1594 &    16.7 &  1.11  & 71.3 &  6.80 &  176 \\
C &   2255 &     15.3 &  1.02 &  63.1 &  3.52  & 110 \\
D &   1625 &     10.7 &  0.711 &  61.5 &   4.24  & 168 \\
E &    927 &    4.00  & 0.267 &  42.7 &  18.4  & 88.1 \\
F &   1694 &    14.7 &  0.978 &  51.8 &  3.22  & 116 \\
G &   2266 &    12.0 &  0.800 &  57.4 &  3.62 &  166 \\
H &   1289 &    4.67 &  0.311 &  67.5 &  44.1  & 97.1 \\
\hline
\hline
AnoDM &  2087 &    17.7 &  1.18 &  70.2  & 4.46 &  228 \\
\hline
    \end{tabular}
    \label{tab:TDEnum}
    \begin{tablenotes}
      \small
    \item Column 2: final mass of the IMBHs, Column 3: the number of the TDEs, Column 4: time-averaged TDE rate, Column 5, 6 and 7: the mean mass, the minimum mass and the maximum mass of the disrupted stars respectively. The first four quantities are averaged using three realizations of the simulations. The latter two quantities are derived from the minimum and maximum mass among the three realizations.
    \end{tablenotes}
  \end{center}
\end{table*}

\begin{figure}
    \centering
    \includegraphics[width=0.5\textwidth]{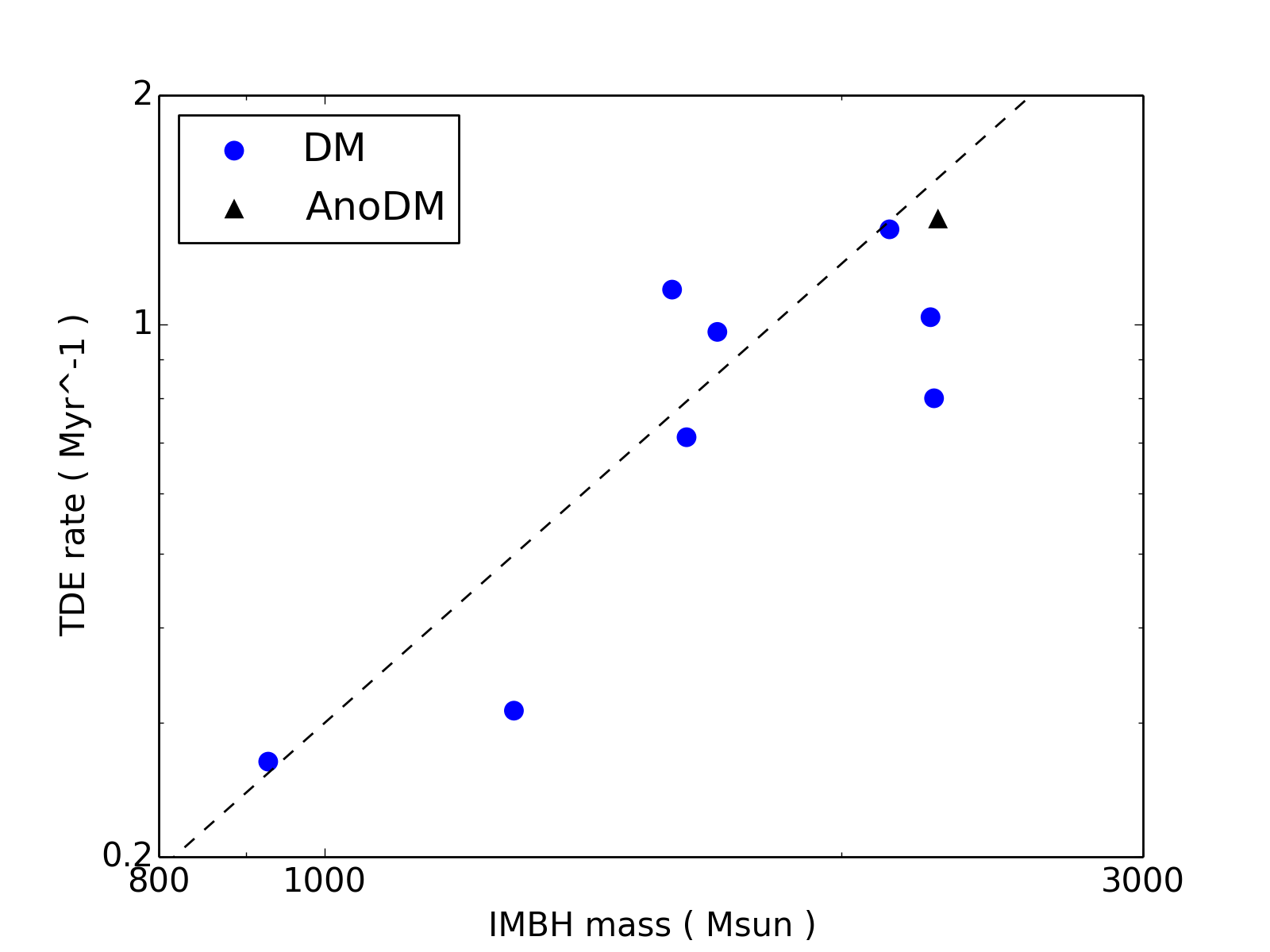}
    \caption{The TDE rates $\dot{N}_{\rm TDE}$ against the IMBH mass $M_{\rm IMBH,f}$. 
    The dashed line is a relation of $\dot{N}_{\rm TDE}=0.3\,\Myr^{-1}\,(M_{\rm IMBH,f}/10^3\,\msun)^2$.
    }
    \label{fig:rate_fit}
\end{figure}

We show the number of TDEs and the TDE rates in \tabref{tab:TDEnum}.
The number of the TDEs ranges from $\sim4$ to $20$, and the TDE rates are $\sim0.3-1.3\,\Myr^{-1}$.
  We show in \figref{fig:rate_fit} that the TDE rates in our simulations correlates
  with the IMBH mass as
\begin{equation}
\dot{N}_{\rm TDE}\sim0.3\,\Myr^{-1}\left(\frac{M_{\rm IMBH,f}}{1000\,\msun}\right)^2.
\label{eq:NdotTDEfit}
\end{equation}
The correlation $\dot{N}_{\rm TDE}\propto M_{\rm IMBH,f}^2$ is suggested in \citet{Baumgardt2004}, 
who estimate the TDE rate as 
\begin{equation}
\dot{N}_{\rm TDE}=
1\,\Myr^{-1}\ 
\rstarnum{0}^{3/5}\ 
\overline{m}_{\rm *,1}^{3/5}\,
\mbhnum{3}^2\,
n_{\rm c,6}^{7/5}\ 
\sigma_{\rm c,1}^{-21/5},
\label{eq:NdotTDE}
\end{equation}
where $n_{\rm c,x}=n_{\rm c}/10^x\,\pc^{-3}$ and $\sigma_{\rm c,x}=\sigma_{\rm c}/10^x\,\kms$
which are the core number density and the core velocity dispersion respectively.
The estimate is based on a loss cone theory presented in \citet{Frank1976}
with calibrations by the results of the N-body simulations of star clusters in \citet{Baumgardt2004}.
We find that our TDE rate can be fitted by the analytical expression (see also \secref{sec:TDE comparison}).

In \tabref{tab:TDEnum}, we show the mean mass, the minimum mass and the maximum mass of the disrupted stars.
The minimum mass ranges $3-44\,\msun$, which is close to
the lowest mass of the IMF of $\sim m_{\rm min}=3\,\msun$ except
for Runs E and H. 
The maximum mass ranges from $88$ to $228\,\msun$.
Note that the massive stars with masses larger than the maximum mass of the IMF $m_{\rm max}=100\,\msun$ are formed by mergers of massive stars with other stars. 
Note also that, though the merged stars would have extended radii for a Kelvin-Helmholtz (KH)
timescale $\sim 1000\,\yr$ \citep{Dale2006,Suzuki2007,Glebbeek2013}, 
we used ZAMS radii for sizes of the merged stars \citep{Tout1996} since 
the KH timescale is short compared to the TDE timescale.
The mean mass of the disrupted stars is $\sim40-70\,\msun$, which is
much higher than the mean mass of the IMF of $\sim 8\,\msun$.
Overall, stars of any mass between $m_{\rm min}$ and $m_{\rm max}$ can be disrupted, and massive stars are preferentially disrupted owing to
mass segregation that occurs rapidly within the dense star clusters.

\begin{figure}
    \centering
    \includegraphics[width=0.5\textwidth]{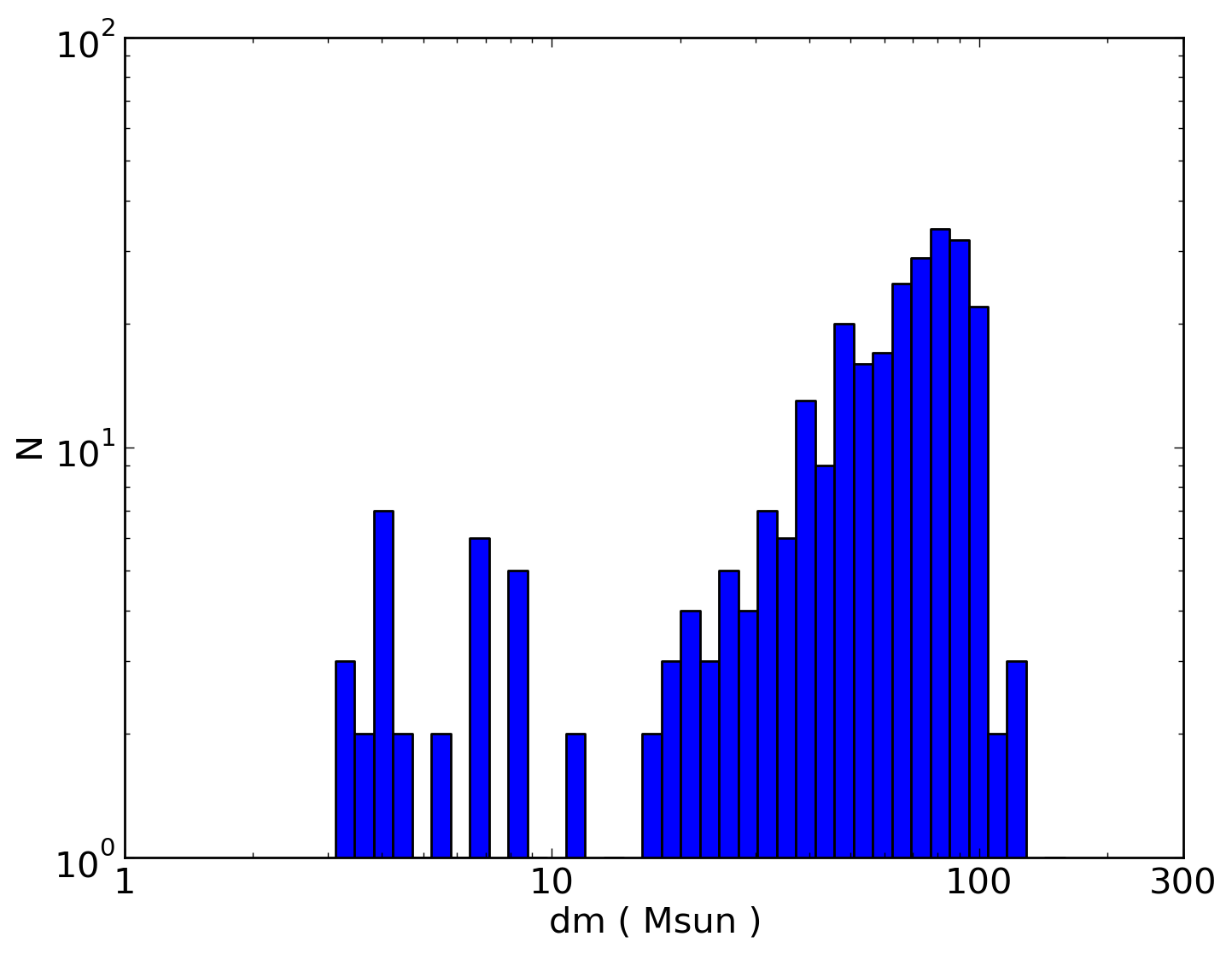}
    \caption{Mass distribution of the disrupted stars. 
    The distribution is generated using all 8 cluster models with DM with three realizations.
    }
    \label{fig:mass distribution}
\end{figure}

We show the mass distribution of the disrupted stars
in \figref{fig:mass distribution} for all our 8 cluster models with DM
with three realizations.
The mass distribution is bimodal with two notable bumps
around $8\,\msun$ and $80\,\msun$.
Stars with high masses are preferentially disrupted
as is already discussed.
The large number of disrupted low-mass stars is due to the
fact that the adopted power-law IMF is weighted
toward low masses.

\begin{figure*}
    \centering
    \includegraphics[width=\textwidth]{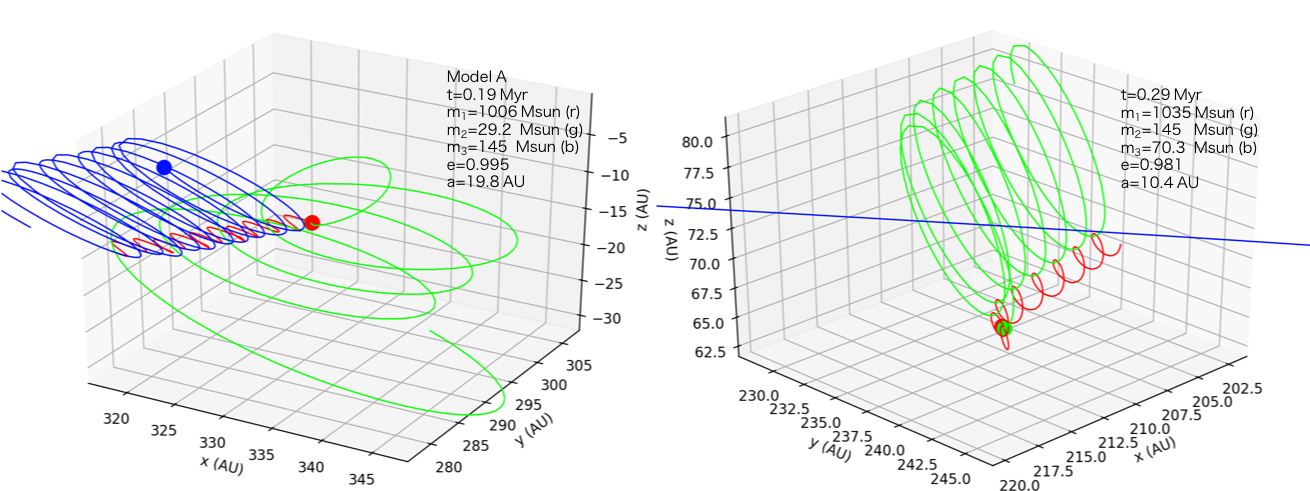}
    \caption{We plot the stellar orbits just before TDEs for two characteristic
      cases. 
      The orbits are taken from the first and second TDEs
      in Run A.
    The points represent the positions of the stars. 
    In the right panel, the tertiary star (blue) passes from left
    to right on a nearly straight orbit before the TDE. 
    }
    \label{fig:orbits}
\end{figure*}

The TDEs are triggered by stellar dynamical interactions around
the cluster centers.
In \figref{fig:orbits}, we show two examples of orbits just before TDEs.
The orbits are taken from the first and second TDEs
in one realization of Run A. 
In the left panel, the IMBH with mass $1006\,\msun$ first forms a binary 
with a massive star with mass $145\,\msun$. 
The third star with mass $29\,\msun$ comes close to the binary and
strongly interacts with it. 
Finally, the latter star collides with the IMBH, leaving the other
massive star, which continues to be a binary component with the IMBH.
In the right panel, the binary is perturbed by a tertiary star
with mass $70.3\,\msun$ which comes close to it. 
The two component stars consequently collide with each other.

\subsection{Global star cluster evolution}
\label{sec:global_ev}
\subsubsection{Model A}
\label{subsec:Model_A}
Before comparing our 8 models, in order to examine the effect of DM on the star
cluster evolution, we first focus on Model A-DM (Model A with DM) and Model AnoDM (Model A without DM).
Hereafter in this section, we refer to a specific realization Run X1 from Model X as Run X, where X is A-DM or AnoDM.

\begin{figure}
    \centering
    \includegraphics[width=0.5\textwidth]{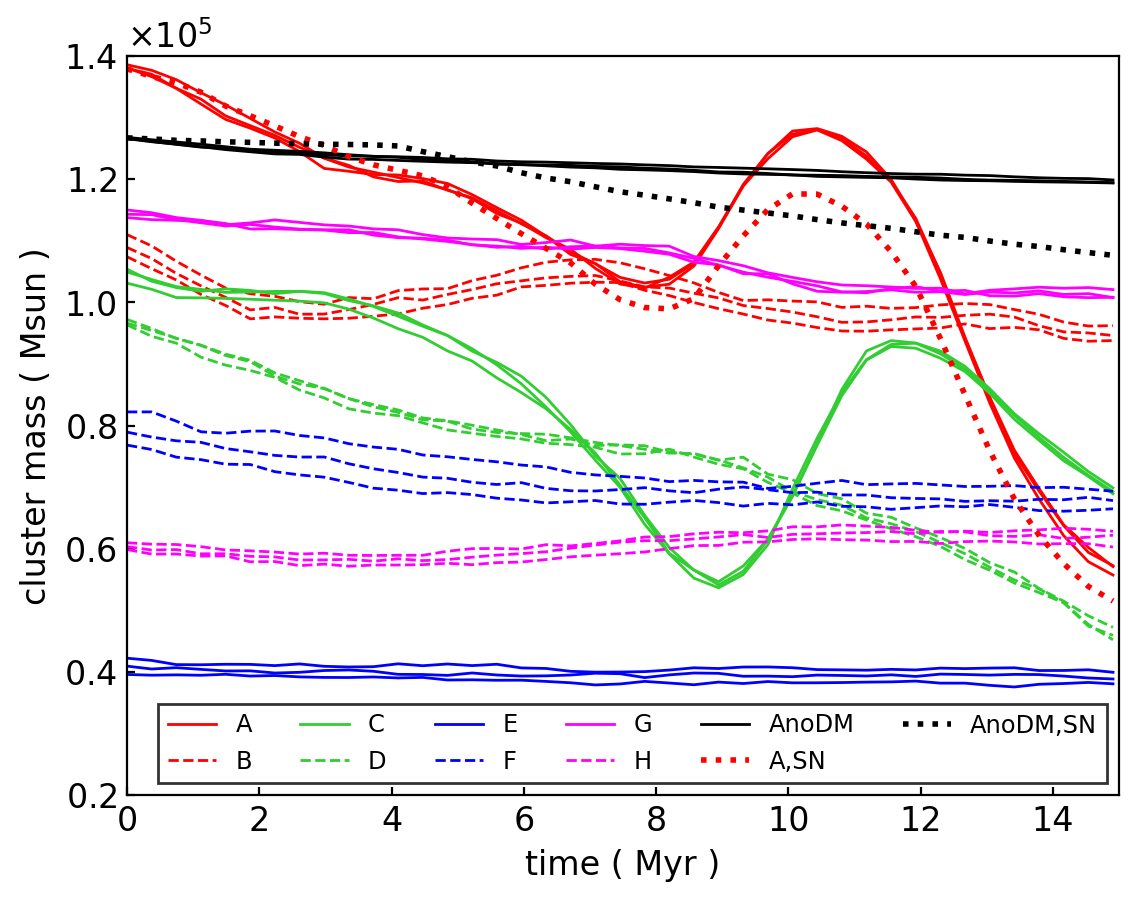}
    \caption{We plot the evolution of the
      bound stellar mass $\mclb$ for each star cluster model.
    We also show the results with a supernova model for Models A-DM and AnoDM 
    (see \secref{sec:length}).
    }
    \label{fig:bmass}
\end{figure}

In \figref{fig:bmass}, we show the time evolution of bound stellar mass $\mclb$ for Model A-DM (red solid lines).
We note again that the `bound stars' mean that the stars are bound
by their own gravitational potential without including the contribution
from the DM component (see Equation \ref{eq:bound cond}).
The bound mass $\mclb$ decreases from $1.4\times10^5\,\msun$ to 
$10^5\,\msun$ until $8\,\Myr$ and then increases to $1.3\times 10^5\,\msun$ until $10\,\Myr$. 
After this point, the mass decreases again
by about 50 percent toward the end of the simulations.

The time variation of $\mclb$ can be explained
by the presence of DM. 
In \figref{fig:bmass}, we also show $\mclb$ for the Model AnoDM
by a black solid line for comparison.
We see little variation in $\mclb$
for the Model AnoDM,
in good contrast to the Model A-DM (red solid lines).

\begin{figure}
    \centering
    \includegraphics[width=0.5\textwidth]{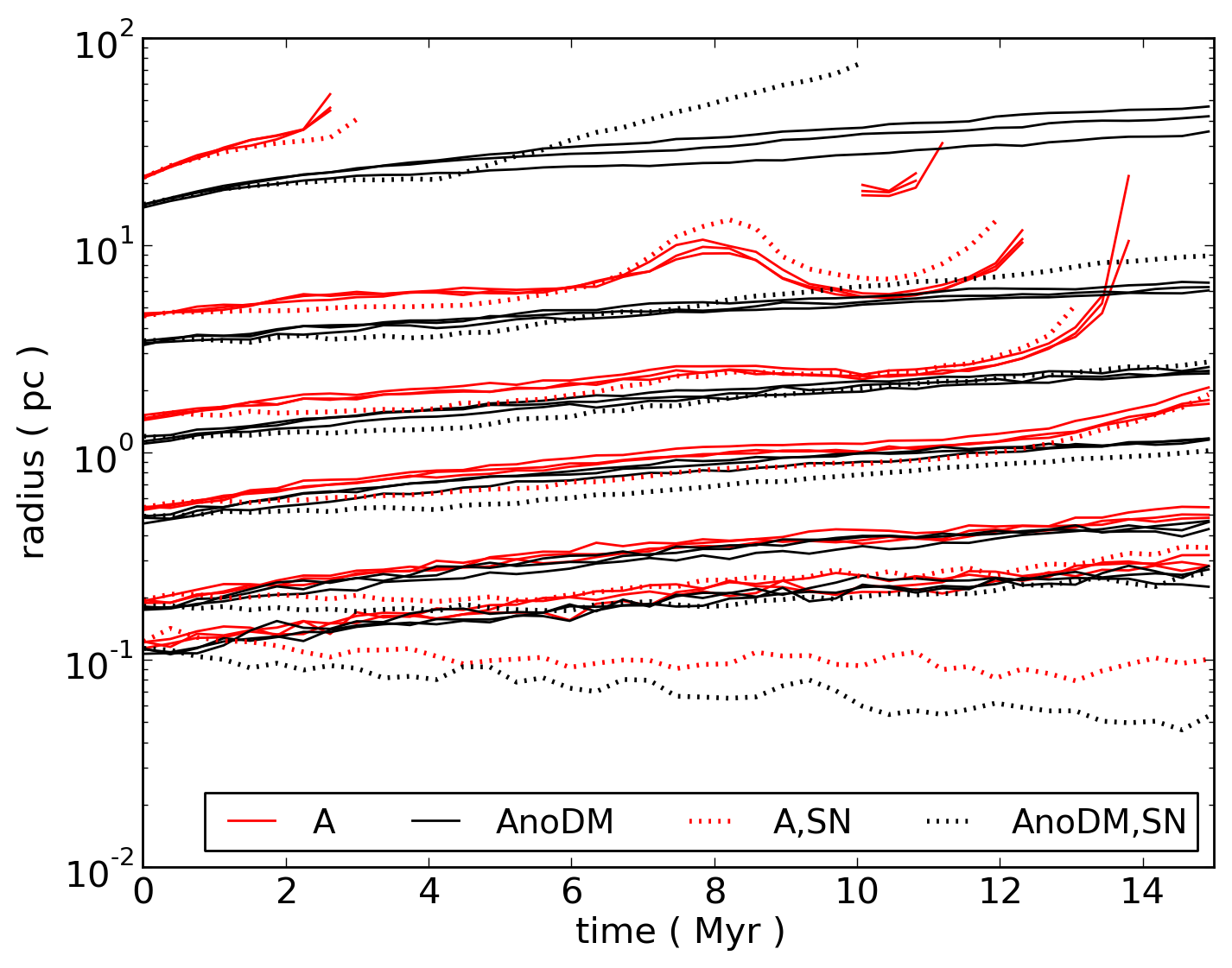}
    \caption{Evolutions of Lagrangian radii for $5, 10, 30, 50, 70$ and $90$\% of the initial bound stellar masses $M_{\rm cl,b0}$ from bottom to top. 
    The red solid lines are from the Model A-DM while the black solid lines are from the Model AnoDM.
    We also show the results with a supernova model for Models A-DM and AnoDM (dotted lines, see \secref{sec:length}).
    }
    \label{fig:r_Lag_dm}
\end{figure}

It is important to identify the cause of variation of
$\mclb$.
In \figref{fig:r_Lag_dm}, we show the Lagrangian radii of
the clusters for $5, 10, 30, 50, 70$ and $90$\% of the initial bound
stellar masses $M_{\rm cl,b0}$. 
The Lagrangian radii for less than $0.3\,M_{\rm cl,b0}$ in the Model A-DM
do not significantly differ from those in the Model AnoDM. 
In contrast, the Lagrangian radii for larger than $0.5\,M_{\rm cl,b0}$
in the Model A-DM increase when $\mclb$ decreases,
while those in the Model AnoDM do not significantly vary.
The expansion indicates that the DM field strips stars from
the outer part of the clusters.

Though the outer part of the stars are stripped by DM, we find that most of 
the stripped stars are still bound by the DM halo in the Model A-DM, 
i.e., $v_{\rm star}^2+\phi_{\rm star}+\phi_{\rm DM}<0$, where $\phi_{\rm DM}$ is a DM potential. 
Furthermore, the DM halo can prevent stars to be completely ejected
from the parent halo even after the stars gain high velocities
after strong binary interactions near the cluster center.

\begin{figure}
    \centering
    \includegraphics[width=0.5\textwidth]{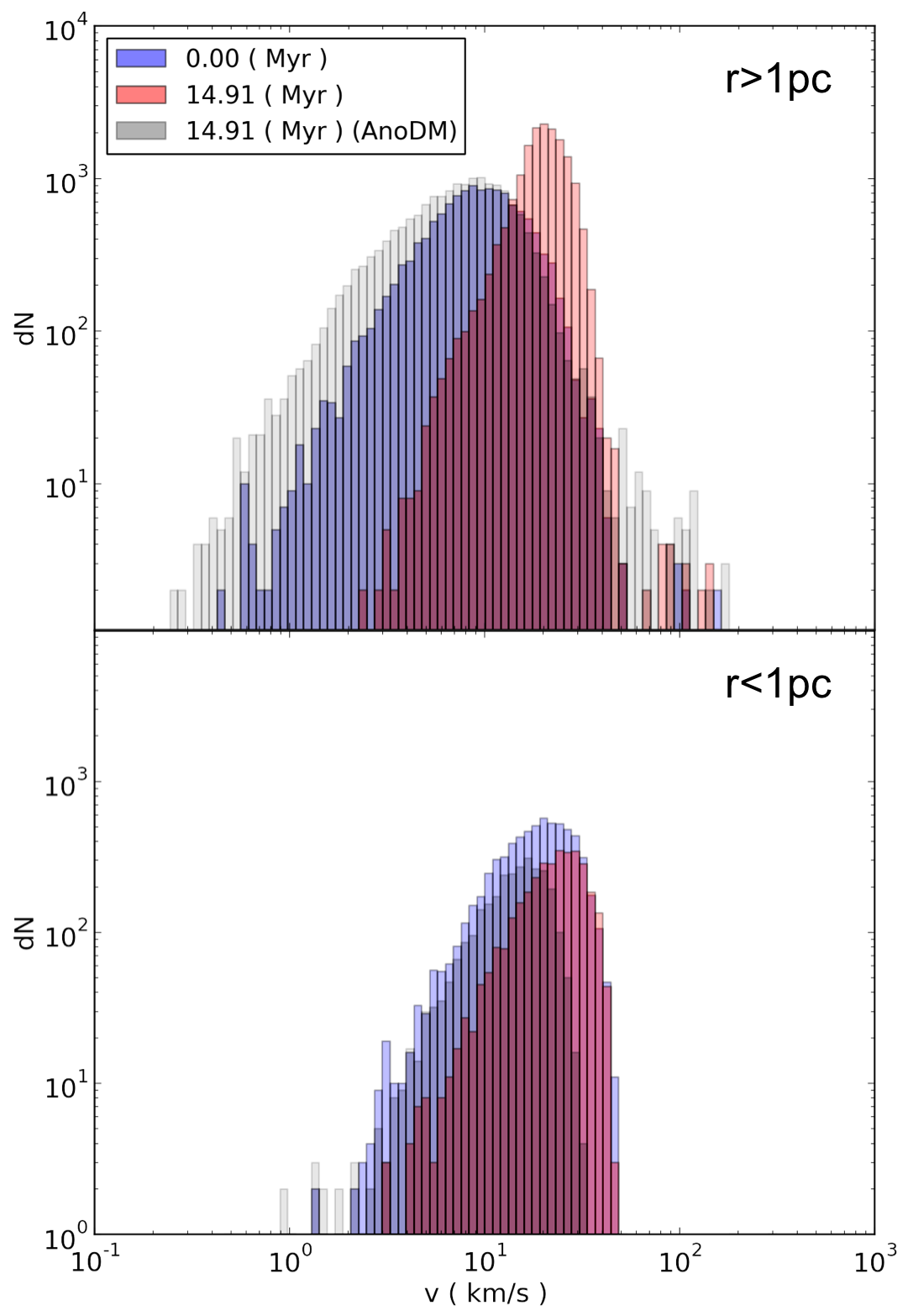}
    \caption{Velocity distributions of the stars for the Run A-DM at $t=0$ (blue) and $15\,\Myr$ (red). The velocity distributions for the Run AnoDM are also shown at $t=15\,\Myr$ (black).
    The top panel shows the distributions for $r>1\,\pc$ while the bottom panel shows those for $r<1\,\pc$. When making the distributions, we adopt width of bins $\Delta\log_{10}(v/\kms)=0.04$.
    }
    \label{fig:hist_vel}
\end{figure}

\begin{figure}
    \centering
    \includegraphics[width=0.5\textwidth]{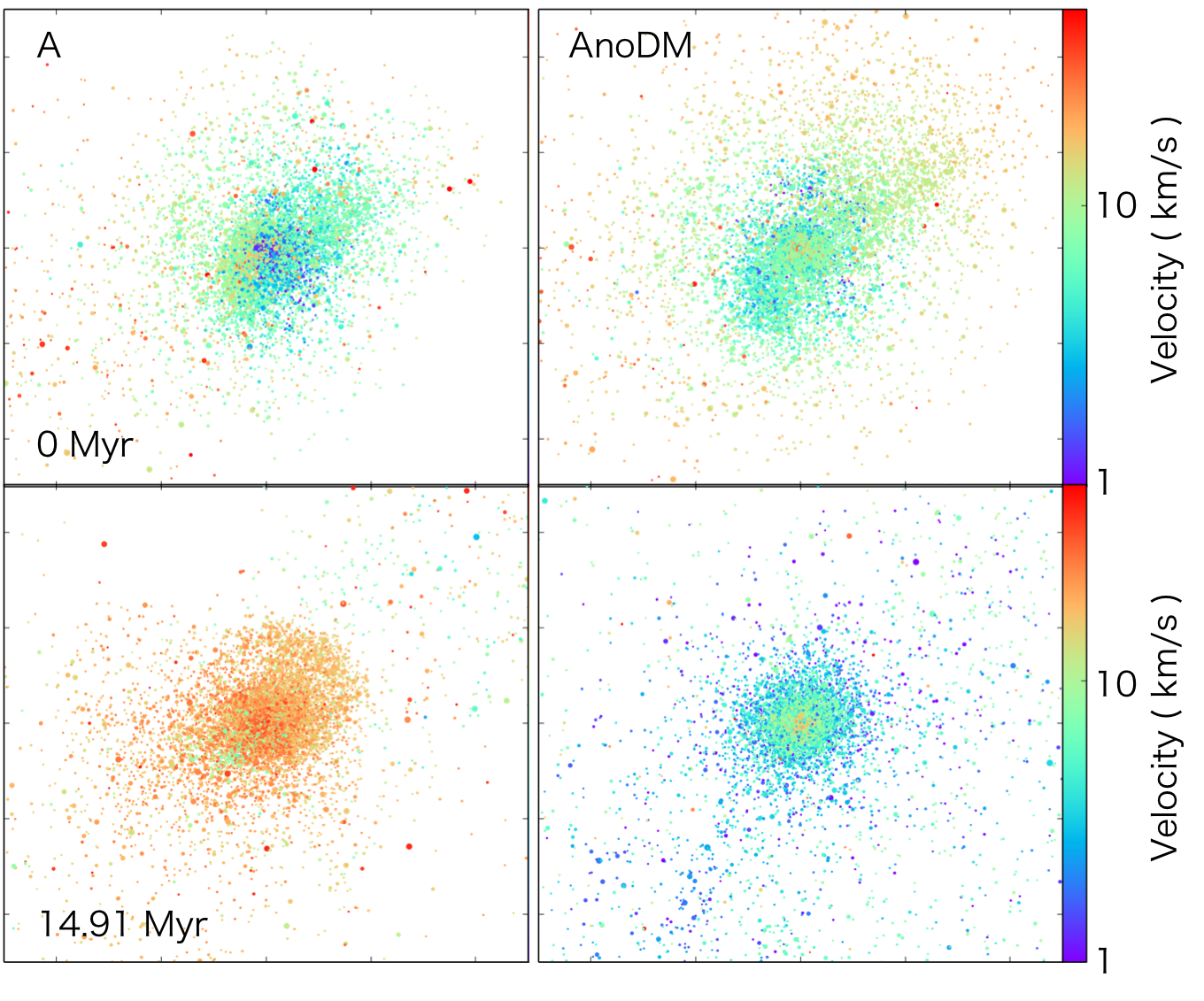}
    \caption{Velocity color map evolution for the Runs A-DM (left) and AnoDM (right) for $t=0\,\Myr$ (top panels) and $t=15\,\Myr$ (bottom panels) in a yz plane. The length of one side is $100\,\pc$.}
    \label{fig:dist_ev}
\end{figure}

The stripping of stars from the outer regions by DM and
the suppression of ejection
can be studied by examining the velocity distributions of the stars.
In \figref{fig:hist_vel}, we show the stellar velocity distribution
for the Run A-DM at 
$t=0$ (blue) and $15\,\Myr$ (red). 
The top panel shows the distributions for $r>1\,\pc$, where $r$ is
the radius from the cluster center.
The distribution of the outer stars shifts with time
toward high velocities, and
the number of high velocity stars with $\gtrsim10\,\kms$ increases
while that of low velocity stars with $\lesssim10\,\kms$ decreases.
The velocity increase is more directly seen from the
velocity color map shown in \figref{fig:dist_ev}.
The number of high velocity stars increases
because of the increase of DM mass in $r\lesssim100\,\pc$. 
According to the virial theorem, stronger gravitational force
increases the stars' velocities.
The peak at $\sim20\,\kms$ in the distribution
at $t=15\,\Myr$ is consistent with the orbital velocity $10\,\kms\lesssim v \lesssim 30\,\kms$ 
for $r_{\rm eq}\simeq10\,\pc \lesssim r \lesssim 100\,\pc$, where $r_{\rm eq}\sim10\,\pc$ is a radius at which the enclosed DM mass is equal to
the enclosed stellar mass.
The velocity enhancement by the increase of the DM mass allows the escape or stripping
of the stars.

The velocity enhancement is not seen in the Run AnoDM
when comparing the blue and black histograms in the top panel
of \figref{fig:hist_vel}
\footnote{Since the velocity distributions are almost the same
  in the Runs A-DM and AnoDM at the beginning of the simulations,
  we do not show the distribution at $t=0\,\Myr$ for the Run AnoDM.} 
  (see also \figref{fig:dist_ev}). 
Interestingly, the distribution spreads with time so that the numbers of
{\it both} low and high velocity stars increase. 
The increase of the low velocity stars is caused by
expansion of the cluster due to two-body relaxation (see black lines in \figref{fig:r_Lag_dm}), whereas
the increase of the high velocity stars with $\gtrsim50\,\kms$ is due to ejections caused by binary interactions in the inner region of the cluster.

Comparing the velocity distributions for the Runs A-DM (red)
and AnoDM (black) at $t=15\,\Myr$, 
we find that high velocity stars with $\gtrsim 50\,\kms$ are less common in the former model. 
This suggests that DM gravity can prevent the ejections of the stars.

In contrast to the outer stars, the presence of DM has little impact on the inner stars.
In the bottom panel of \figref{fig:hist_vel}, we show the velocity distributions for $r<1\,\pc$. 
The distribution in the Run A-DM does not significantly change with time;
indeed, the distributions are similar
between the Runs A-DM and AnoDM at $t=15\,\Myr$.

\begin{figure}
    \centering
    \includegraphics[width=0.5\textwidth]{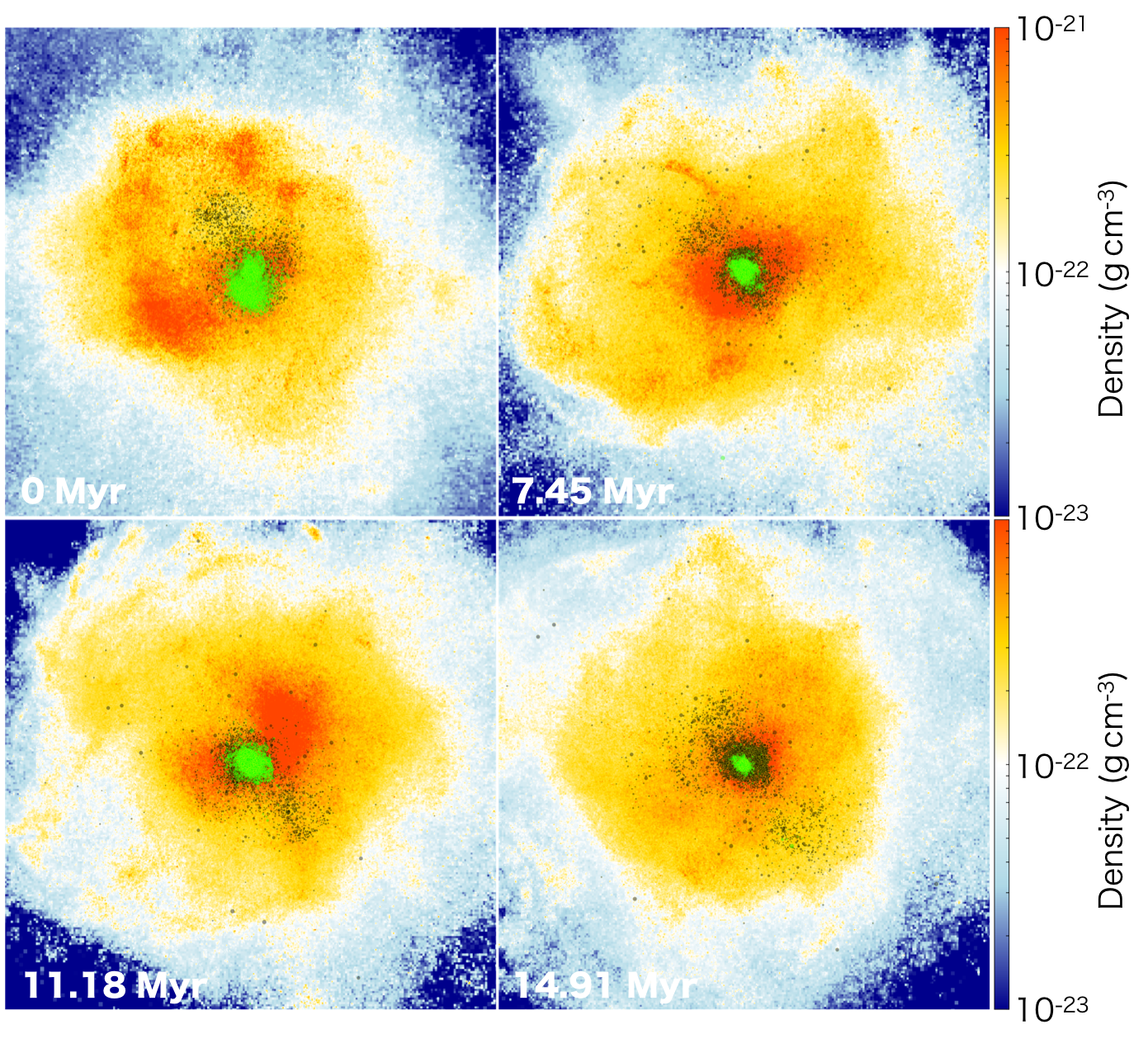}
    \caption{Snapshots of DM density (color) and stellar distributions (dots) for the Run A-DM in a zx plane.
    The green and black dots denote bound and unbound stars respectively.
    The length of one side is $400\,\pc$.
    }
    \label{fig:pos_A_evolution}
\end{figure}

We find that the DM motion relative to the star cluster
causes the oscillatory behavior of the bound mass evolution for the Model A-DM in \figref{fig:bmass}.
In \figref{fig:pos_A_evolution}, we show snapshots of DM density (color) and stellar distributions (dots) at four epochs.
A high density DM clump which is located at about $100\,\pc$
lower-left from the center at $t=0\,\Myr$ moves towards
the upper right direction.
At $t=7.45\,\Myr$, the clump temporarily merges with the high
density region where the star cluster resides,
enhancing the DM density at $r\lesssim50\,\pc$. 
Then the bound stellar mass $\mclb$ (\figref{fig:bmass}) decreases
with the increase of the stellar velocity.
The DM clump then passes through the central region at $t=11.18\,\Myr$,
and $\mclb$ increases again.
Finally, the clump merges with the central region,
and begins stripping the stars from the cluster.

\begin{figure}
    \centering
    \includegraphics[width=0.5\textwidth]{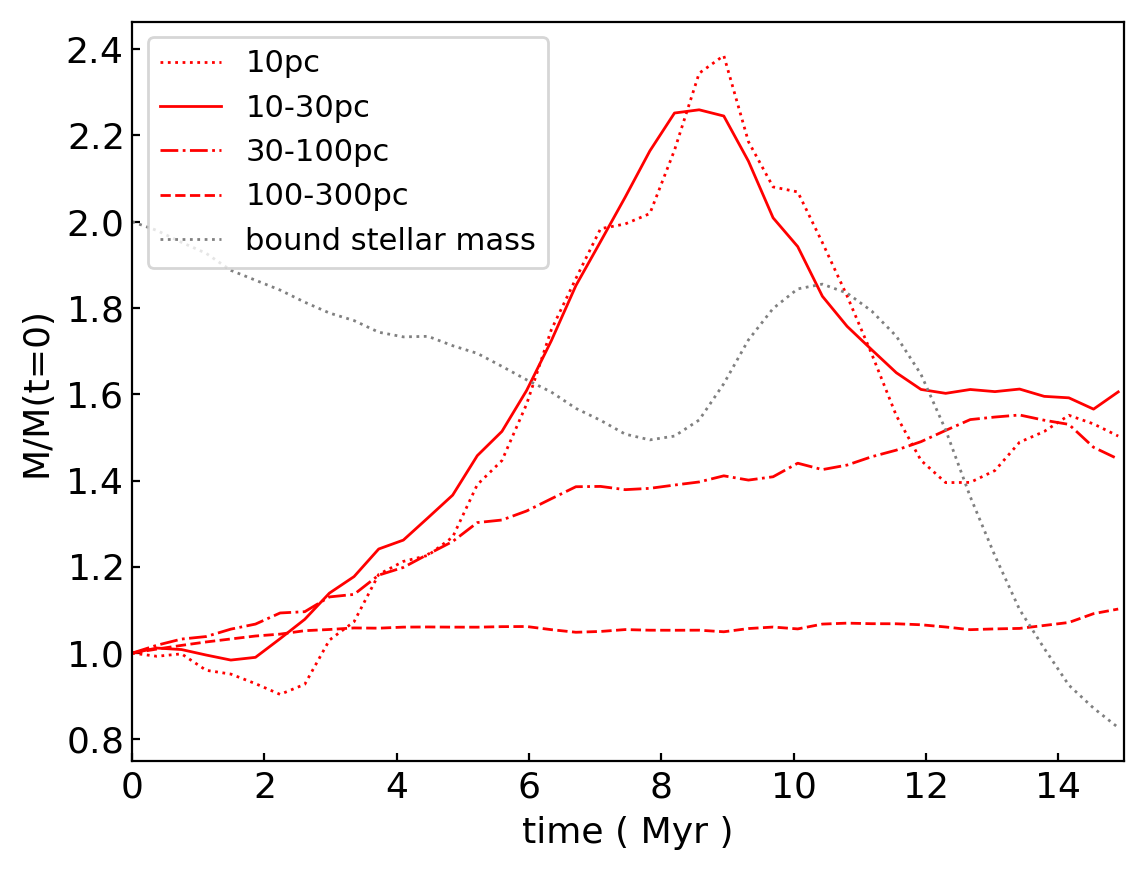}
    \caption{Evolution of DM masses ($M_{\rm DM}$) normalized at the initial masses in regions of  
    $r<10\,\pc, 10-30\,\pc, 30-100\,\pc$ and $100-300\,\pc$. 
    The initial masses in the regions are $M_{\rm DM}(t=0)=6.3\times10^4, 8.4\times10^5, 1.1\times10^7$ 
    and $2.8\times10^7\,\msun$, respectively.
    We also plot the corresponding bound stellar mass evolution $2\mclb/M_{\rm cl,b0}$, where a factor 2 is introduced to clearly show the oscillation (see \figref{fig:bmass}).}
    \label{fig:em_dm_norm}
\end{figure}

The first and the second drops of $\mclb$ in \figref{fig:bmass} are caused by a series of processes associated with DM motions.
To see this, in \figref{fig:em_dm_norm} we show the evolution
of normalized DM mass $M_{\rm DM}/M_{\rm DM}(t=0)$ at several regions $r<10\,\pc, 10-30\,\pc, 30-100\,\pc$ and $100-300\,\pc$.
For $\lesssim8\,\Myr$ during which the first drop of $\mclb$ occurs, the DM mass in the inner regions 
$\lesssim30\,\pc$ increases by a factor of $2$. 
The mass increase indicates that the DM falls toward the central region and deepens
the potential well, causing the first drop.
At $\gtrsim11\,\Myr$ during which the second drop occurs, the inner DM mass is roughly constant and instead the outer mass in $30-100\,\pc$ continues to increase.
As the outer DM mass increases, stars in the outer regions are accelerated
and get unbound, causing the second drop.

\begin{figure}
    \centering
    \includegraphics[width=0.5\textwidth]{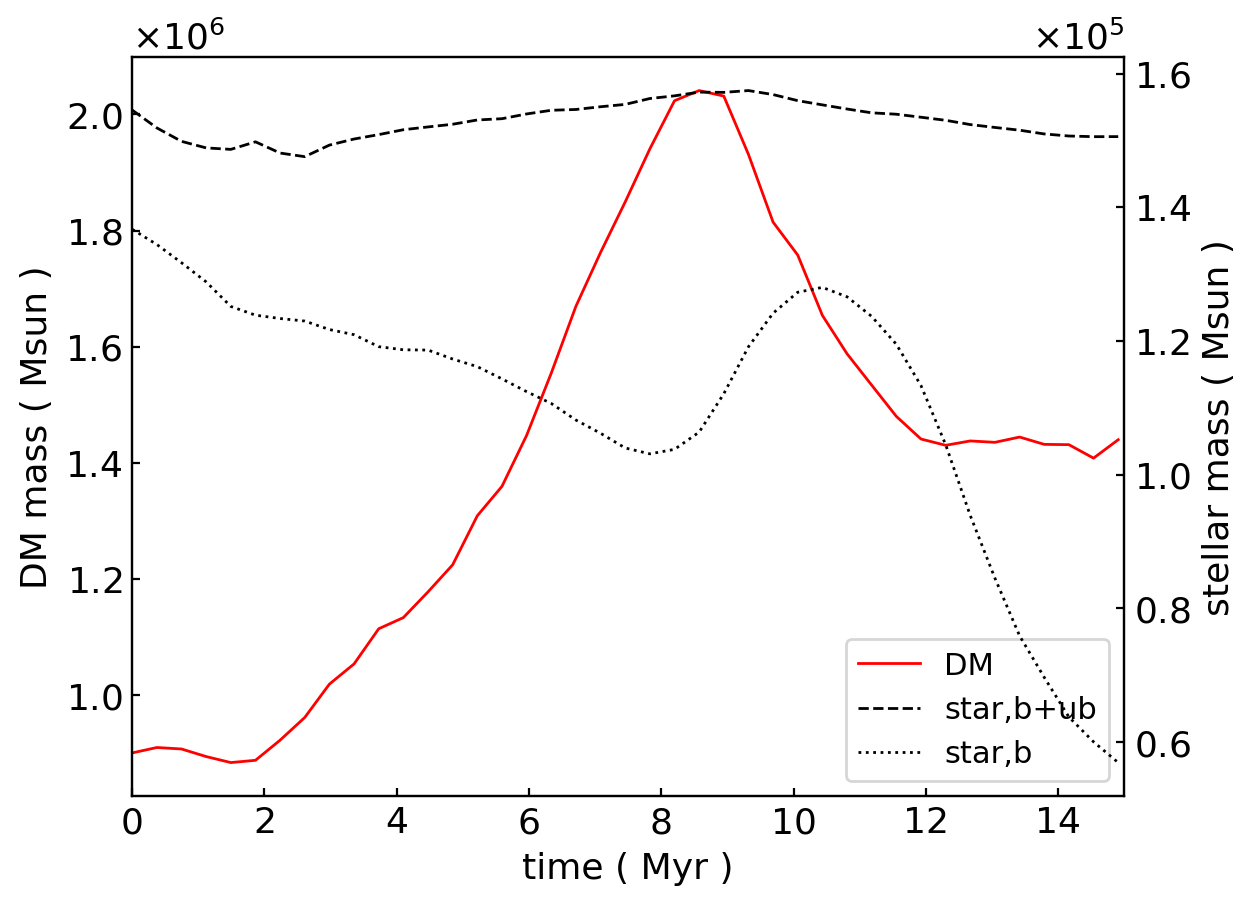}
    \caption{Evolutions of DM mass (red solid), stellar mass including both bound and unbound stars (black dashed) 
    and bound stellar mass (black dotted) within $30\,\pc$ from the center. 
    }
    \label{fig:em_dm}
\end{figure}

To further examine the dynamical reaction of the stars when
the DM clump falls toward the central 
region, we compare the evolution of DM mass and stellar mass within $30\,\pc$ in \figref{fig:em_dm}.
At $t\lesssim8\,\Myr$, the DM mass increases since the DM streams
from the outer region to within $30\,\pc$. 
Later, the DM mass decreases until $t\sim12\,\Myr$ and
then remains roughly constant with $\sim1.4\times10^6\,\msun$.
The evolution of the stellar mass including both bound and unbound stars
(dashed line) traces that of the DM: the stellar mass also increases 
for $t\lesssim8\,\Myr$ due to the increase of the DM mass but
decreases afterwards.
Note that the opposite is true for bound stellar mass evolution (dotted line): the stellar mass decreases as the DM mass increases due to the velocity enhancement and stripping.

\begin{figure}
    \centering
    \includegraphics[width=0.5\textwidth]{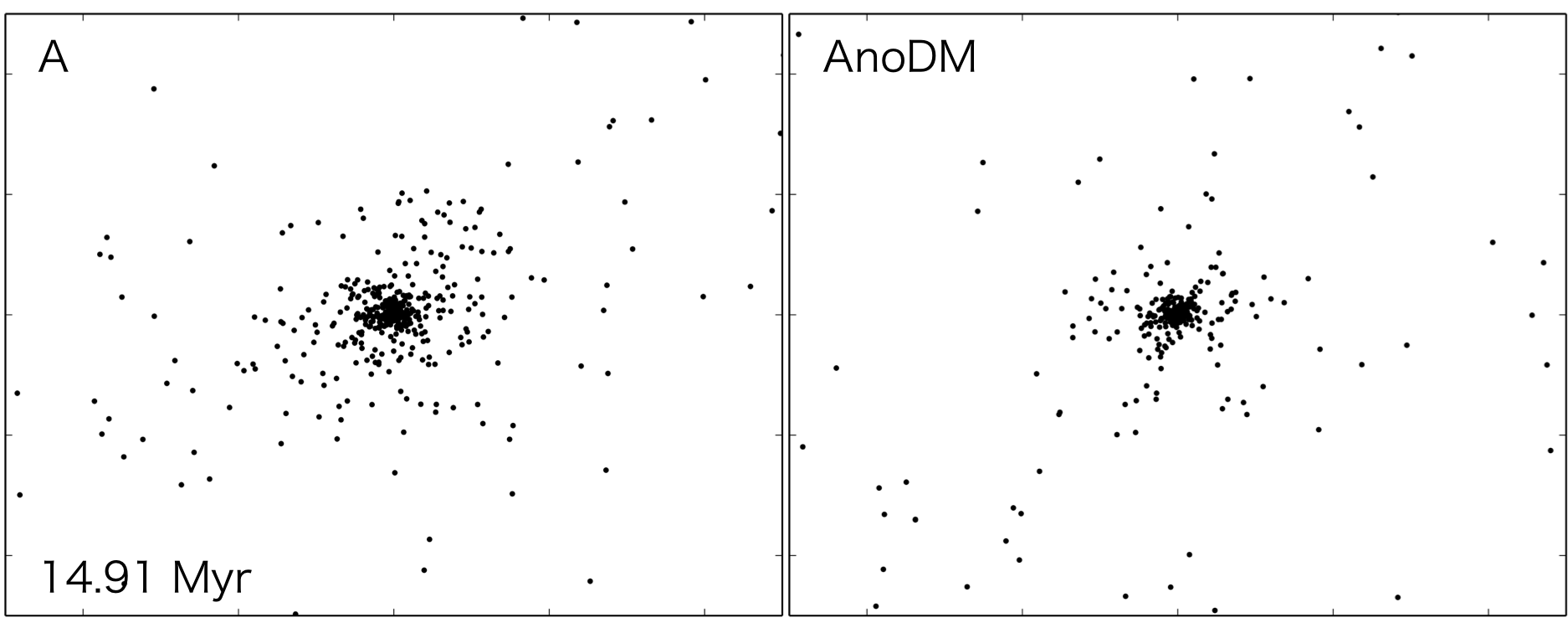}
    \caption{The distributions of massive stars with mass $m>30\,\msun$
      for Run A-DM (left) and AnoDM (right) in a yz plane.
      The length of one side is $100\,\pc$. 
    The number of massive stars with $m>30\,\msun$ is $690$ and $688$, respectively.
    }
    \label{fig:pos_massive}
\end{figure}

The DM motions can also affect the TDE rate. 
To show this, we follow a longer time evolution
of Models A-DM and AnoDM to $\sim22.5\,\Myr$. 
In \tabref{tab:TDEtime}, we show the TDE rates before and after the DM falls in the central region,
$t<15\,\Myr$ and $t>15\,\Myr$.
Before the DM density increases, the TDE rate is higher in Model A-DM.
Afterwards, however, the rate becomes {\it lower} than in Model AnoDM.
The faster decrease of the TDE rate is attributed to the increase of 
DM density that deepens the total gravitational potential,
and to stripping of massive stars that weakens mass segregation
within the star cluster.
\figref{fig:pos_massive} shows clearly an extended stellar distribution of massive stars in Run A-DM.

\begin{table}
  \begin{center}
    \caption{TDE rates for $t<15\,\Myr$ and $t>15\,\Myr$ in Models A-DM and AnoDM. 
    The values are averaged using three realizations of the simulations.}
      \begin{tabular}{lcc} \hline
      Model &      $\dot{N}_{\rm TDE}(t<15\,\Myr)$ & $\dot{N}_{\rm TDE}(t>15\,\Myr)$ \\
                 &    ($\Myr^{-1}$) & ($\Myr^{-1}$)  \\
                 \hline \hline
A &  1.33 & 0.311 \\
AnoDM &  1.18 & 0.533  \\
\hline
    \end{tabular}
    \label{tab:TDEtime}
  \end{center}
\end{table}

\subsubsection{Comparison among the models}
There are notable differences in the time evolution
of $\mclb$ between our models (\figref{fig:bmass}).
In Model A and C, $M_{\rm b}$ oscillate with time,
whereas $M_{\rm b}$ decreases monotonically 
from $\sim10^5\,\msun$ to $5\times10^4\,\msun$ in Model G.
The other models do not show significant variation of $M_{\rm b}$
and exhibit little decrease or no change.

The different behaviors are attributed to the details of the DM motions. 
For example, in our Model D, a DM clump gradually falls in the central region
, leading to the decrease of $\mclb$ (\figref{fig:bmass})
as is also seen in Model A.
Contrastingly, in Model G, the central DM density does not vary significantly
and $\mclb$ changes little.
Overall, we conclude that the degree of stellar stripping
and the behaviour of star clusters sensitively depend on the details
of the DM dynamics.


\section{Discussion}
\label{sec:discussion}

\subsection{Comparison of TDE rates}
\label{sec:TDE comparison}
We estimate the typical TDE rate to be $\sim0.1-1\,\Myr^{-1}$ for our clusters
with mass $M_{\rm cl}\sim10^5\,\msun$ (\figref{fig:rate_fit}).
Using the box size of the cosmological simulations $10h^{-1}\,\Mpc$ and the number of
the clusters in the box, the ``cosmic'' TDE rate is inferred, very roughly,
to be $\sim1-10h^3\,\Gpc^{-3}\,\yr^{-1}$.

Although there is no TDE observation at $z\sim10-20$ which can be directly compared to our estimate, 
it is intriguing to compare it with a local observation in order to gain an insight of difference in properties between the primordial and local clusters.
Recently, \citet{Lin2018} find a X-ray outburst 3XMM J215022.4-055108 which might originate
from a tidal disruption of a star by an IMBH of a few $10^4\,\msun$ in a massive star cluster with mass $\sim10^7\,\msun$. 
They estimate that a TDE rate is $\sim10\,\Gpc^{-3}\,\yr^{-1}$.
If we scale the TDE rate as in Equation \ref{eq:NdotTDEfit} to BH mass with a few $10^3\,\msun$ which is comparable to our IMBH masses, 
the TDE rate would be $\sim0.1\,\Gpc^{-3}\,\yr^{-1}$.
This value is close to our estimate, implying that the primordial and local clusters have some similar properties.
For example, cluster core density and cluster velocity dispersion would be similar, 
otherwise TDE rates would be significantly different considering that the rates have strong dependencies on those properties (see Equation \ref{eq:NdotTDE}).
Although there are a number of differences in, e.g., the number of stars within a cluster, mass function of stars and stellar populations, 
the comparison of the TDE rates can be a probe to explore differences in the distant and local cluster properties.
Previous theoretical studies also estimate the rate of stellar TDEs by IMBHs 
(\citealt{Baumgardt2004}; \citealt{Stone2017}; \citealt{Chen2018,Fragione2018}).
For example, \citet{Baumgardt2004} derived a TDE rate of Equation \eqref{eq:NdotTDE} by using 
a loss cone theory.
As we show in \figref{fig:rate_fit}, the TDE rate estimated from the equation is consistent 
with ours.
\citet{Stone2017} also derived a TDE rate from their results for mass growth histories of BHs (see their figure 8). 
From the figure, we estimate the TDE rate as $0.1-1\,\Myr^{-1}$ for a cluster that has a core radius 
$0.1\,\pc$ and an IMBH with mass $10^3\,\msun$, assuming that the masses of tidally disrupted stars 
are $10-100\,\msun$.
This TDE rate is also consistent with our result.

Because we do not consider the processes like tidal captures \citep[e.g.,][]{Baumgardt2006},
which could enhance the TDE rate, our derived rate in the above could be conservative.
We also note that the deviation of the initial conditions from an isotropic distribution could
cause suppression of stellar flux to the central IMBH \citep{Lezhnin2015}.

\subsection{Implications to `halo stars'}
We have shown that stars in the outer part of a star cluster can be stripped
by the presence of DM.
The stripped stars are still bound within the DM halo, and then the stars would
become `halo stars' which spread out beyond the cluster tidal radius, similarly
to the observed stellar populations in local star clusters \citep{Marino2014}.
\citet{Penarrubia2017} use analytic models and numerical experiments
to show that the presence of DM helps forming a stellar halo.
With the DM, the density profile of stars becomes shallower at large distances 
and the velocity dispersion of the stars is increased.

\subsection{Evaporation of star clusters and the fates of IMBHs}
\label{sec:evaporation}
An isolated star cluster can be evaporated
by the two-body relaxation process in a typical timescale of $10-100 t_{\rm rh}$ \citep{Spitzer1958,Johnstone1993}, with the exact timescale depending on the stellar mass distribution.
If there are tidal shocks or tidal force from a galaxy, the evaporation timescale
can be as short as or shorter than $10t_{\rm rh}$ \citep{Gnedin1997,PortegiesZwart2001}.

The cluster evaporation can be also accelerated by the motion of the DM:
DM strips stars from a cluster when it increases the effective mass density
within the cluster and accordingly increases
the stellar velocity dispersion. 
The evaporation would eventually cause destruction of the star cluster.
Although we have shown that the inner regions of the star clusters are not significantly affected
by the DM, it is likely that 
the inner regions would be also dissolved if the evolution is followed over a much
longer time, since the inner region 
$r\lesssim 1\,\pc$ actually expands gradually (see \figref{fig:r_Lag_dm}).

When a star cluster evaporates, the central IMBHs can no longer grow
by collisions of stars due to scarcity of intruding stars.
For the IMBHs to grow and become supermassive, a further external supply of mass by gas accretion or stellar/BH mergers is necessary.
For example, if the host halo merges with other halo(s) or galaxies,
gas will fuel the central IMBHs,
and finally BH mergers will increase the mass \citep[e.g.,][]{Kauffmann2000}. 
Large-scale cosmic gas flows can also fuel the BH \citep[e.g.,][]{Dekel2006, Bellovary2013}.

IMBHs formed at the cluster centers would also continue floating
in the sea of stars which are not bound
by themselves but are still {\it loosely} bound by the DM halo. 
Even in this case the stars and IMBHs could be incorporated
in other galaxies during the cosmic evolutions.
The origin of some observed IMBH candidates (\citealt{Maillard2004,Schodel2005};\citealt{Chilingarian2018})
may be explained by our IMBH formation scenario.

It is interesting to study the ratio of the BH mass to the cluster/bulge mass,
  which is determined when the BH mass growth ceases.
  \figref{fig:Mcl-mbh} shows that the IMBH mass is $1-2$\% of the star cluster mass. 
Our result is consistent with those obtained
by a theoretical work \citep{PortegiesZwart2002}, meaning that DM does not significantly affect the shape of the relation.
Our result is also consistent with those in local observations (\citealt{Lutzgendorf2013}, see also figure 6 of \citealt{Sakurai2017}), indicating that the observed candidate IMBHs could have been formed 
by the runaway stellar collision scenario.
When we extrapolate our results to SMBH regimes
and compare them with the well-known SMBH mass-bulge mass
relation \citep{Magorrian1998,Merritt2001}, we overestimate the BH mass by an order of magnitude.
During evolution from IMBHs to SMBHs, substantial mergers and accretion could have occurred to shape the Magorrian relation.

\subsection{Caveats of the simulations}
\subsubsection{Impacts of IMF}
The adopted minimum mass of the Salpeter IMF $m_{\rm min}=3\,\msun$ in our simulations is larger than the typical value $\sim0.1\,\msun$. 
Sampling the IMF with a smaller $m_{\rm min}$ could alter our results\footnote{
In SYFH17, we showed that decreasing $m_{\rm min}$ to $1\,\msun$ does not significantly alter results in a simulation without DM (see their model Amin in table 3). 
Decreasing $m_{\rm min}$ to a much smaller value or including DM, however, could alter the results.
}.
For example, when fixing a total cluster mass, decreasing $m_{\rm min}$ decreases the average stellar mass in the star clusters, increases a total number of stars, and decreases a number of massive stars.
Then, the average mass of stars which are tidally disrupted by IMBHs would decrease. 
On the other hand, the TDE rates might increase because the number of low mass stars which sink toward the IMBH due to mass segregation increases \citep[see equation 15 in][]{Fujii2008}.
The resulting final IMBH mass would be altered depending on relative significance of these effects. In \citet{Gurkan2004}, the IMBH mass changed only by a factor of 2, even though they assumed various mass functions.
In addition, by decreasing $m_{\rm min}$, a number of low mass stars which reside in an outer part of the cluster would increase. 
These low mass stars would be more easily accelerated or decelerated by the motion of DM. 
Hence, the evolution of the bound stellar mass in the cluster (\figref{fig:bmass}) would be more susceptible to the DM motion.

To verify the effects of adopting a smaller $m_{\rm min}$, it is necessary to perform more computationally expensive simulations.
Firstly, decreasing $m_{\rm min}$ with fixing the total stellar mass means increasing the number of stars in the cluster.
Secondary, by decreasing $m_{\rm min}$, we need to also decrease the DM particle mass $1.87\,\msun$ to a lower value in order to avoid numerical effects.

We also note that in primordial star clusters the IMF is not necessarily the Salpeter IMF.
Performing simulations with changing the IMFs are intriguing future works for revealing overall impacts of IMFs.

\subsubsection{Stellar evolution}
\label{sec:length}
We follow the cluster evolution for a longer time of $15\,\Myr$ than the lifetimes of massive stars.
For example, a star with mass $\sim15\,\msun$ with solar metallicity
has a main-sequence (MS) lifetime of $\sim10\,\Myr$ \citep{Buzzoni2002}. 
The lifetimes are shorter for stars with lower metallicity \citep{Schaerer2002}.

At the end of their lives, stars with mass between $8$ and $40\,\msun$ are
thought to explode as supernovae and lose a substantial amount of mass 
\citep{Heger2003}.
In order to test if and how the mass loss by the supernovae 
affects the global evolution of the clusters, we perform
additional simulations with a model of supernovae \citep{Hurley2000} 
for the model A with and without DM
\footnote{Although we consider a low metallicity environment in the present paper, 
we use a model for the solar metallicity to facilitate the test simulations.}.
In the simulations, we convert stars to compact remnants at the end of their lifetimes. 
We also model mass loss for stars $\lesssim25\,\msun$ 
\footnote{The threshold mass $25\,\msun$ is rather arbitrary, but changing this value to higher mass does not affect the following argument.}, 
whereas more massive stars are assumed to collapse to leave remnant BHs.
We suppress close encounters of stars/compact objects by increasing 
the softening length for the simulation particles from $0$ to $\sim0.0078\,\pc$ in order 
to prevent formation of extremely hard binaries and to make computation feasible. 
In this situation, stellar collisions/TDEs did not occur, and IMBHs did not grow.
Also in the test runs, we do not consider possible gravitational wave events by BH mergers.
In \figref{fig:bmass}, we show the evolution of bound stellar mass $\mclb$ for our
test simulations with dotted lines.
The supernovae accelerate expansion of the clusters and evaporation of stars from the clusters.
In \figref{fig:r_Lag_dm}, we show the evolutions of Lagrangian radii for the test simulations.
While the Lagrangian radii for $>0.7\mclbz$ expand faster than in the simulations without the supernova model and the overall cluster expansion is confirmed, 
the Lagrangian radii for $<0.1\mclbz$ shrink faster.
The outer radii expand and the inner radii shrink because 
\begin{enumerate}
\item massive stars which do not lose mass after their deaths preferentially reside in the center of the clusters while low mass stars which lose mass reside in the outer parts of the clusters,
\item due to energy equipartition, stars are supplied to the outer parts from the central part of the clusters and 
\item the central part of the clusters shrinks in order to become virialized.
\end{enumerate}
The supernovae would not stop stellar TDEs and the growth of the IMBHs since the central density of the clusters increases.

Although stars with masses between $40\,\msun$ and $60\,\msun$ would directly collapse to BHs, 
massive stars with $\gtrsim60\,\msun$ could decrease their masses by pulsational pair instability supernovae (PPISNe) and pair instability supernovae (PISN) in $15\,\Myr$ \citep[e.g.,][]{Belczynski2016, Spera2017}.
Mass loss by PISNe would have minimal effects on our results since there are few stars which have masses $\gtrsim100\,\msun$ and can explode as PISNe.
Mass loss by PPISNe, on the other hand, could alter the cluster evolutions. 
By PPISNe, stars with masses $\sim60-100\,\msun$ can lose more than $50$ per cent of their masses by the end of their lifetimes, which suppresses mass segregation and enhances the expansion of the clusters.
The TDE rates would consequently drop.

After supernovae, stellar mass BHs are left and they could merge with IMBHs in the star clusters.
The merger product would then get a recoil velocity which can alter the results. 
If we assume a BH mass $100\,\msun$ and an IMBH mass $1000\,\msun$, the recoil velocity is 
$\sim30\,\kms$ for the Schwarzschild BHs (figure 3 of \citealt{Fitchett1984}).
The recoil velocity would be higher if the merging binary BHs have equal mass or if they have spins \citep{Gonzalez2007}.
Although the escape velocity of the clusters is $\sim 30\,\kms$, the IMBHs may, nevertheless, remain in the star clusters since the escape velocity could be higher due to the presence of the DM halo and the recoil velocity could be lower than the latter escape velocity
\citep[see][]{Morawski2018}.

There is another complication, actually a simplification in our model,
related to the evolution timescales.
Star formation likely lasts for a finite period of time. 
For example, the radiation hydrodynamics simulations of
\citet{Kimm2016} show that star cluster formation within an atomic-cooling halo
lasts for $\sim 10\,\Myr$. 
To fully examine the effect of the stellar evolutions on the star cluster evolutions,
it is necessary to consider self-consistent processes
from star formation to the cluster dynamical evolution.

\subsubsection{Effects of gas}
Our simulations do not incorporate a diffuse gas component that
may remain after the star cluster formation.
With the presence of gas, IMBHs in the star clusters can increase mass by gas accretion \citep[e.g.,][]{Kawakatu2005,Leigh2013}.
\citet{Leigh2013} argue that gas accretion onto BHs in star clusters occurs rapidly and helps
the BH growth.
Although they consider larger star clusters than ours, it is worth examining whether IMBH growth is promoted by the accretion.

The diffuse gas can also alter the stellar dymamics.
\citet{Lupi2014} use a semi-analytic model
and show that the gas inflow can help stellar BHs to avoid ejections
from the cluster and can also help runaway collisions by deepening the potential well.
\citet{Sills2018} argue that the diffuse gas remaining in star clusters
can decrease the central density of the star clusters if the gas mass is $10$ times larger than the stellar mass.

In our star cluster models, the gas inflow from the outer part of the parent halos
may actually promote TDEs.
However, since the mass of the gas that we removed when generating the initial
conditions is $10-20\,M_{\rm cl}$ 
\citep[table 1 of][]{Sakurai2017}, it could decrease the density of the cluster and TDE rates. 
Examining either of the two effects has larger effect is an intriguing future work.

\subsubsection{Mass increase rates of the IMBHs}
\label{sec:mass increase}
We may overestimate the mass increase of the IMBHs 
by assuming that a tidally disrupted star is fully swallowed.
If we assume the mass
of the disrupted star $\mstar=100\,\msun$ and $\mbh=1000\,\msun$
with $\beta=1$, the star can be fully bound after the disruption
only when its orbit
has a small eccentricity $e\lesssim0.07$
\citep[see equation 20 of][]{Hayasaki2013}.
In our simulations, the eccentricities of the disrupted stars are often $e\sim1$. 
In such parabolic-orbit cases, about a half of the disrupted ``debris'' is actually
ejected outward, and only another half is accreted on to the IMBH \citep{Rees1988}.
Considering the possible partial ejection, the IMBH growth
derived in our simulations may be the upper limits.
Despite the overestimation, our result that the IMBH grows moderately by the TDEs does not change.

There are some other physical mechanisms that can affect our results. 
For example, an accretion disk can be
formed around an IMBH after TDE(s) \citep[e.g.,]{Shen2014}.
The disk, if survives for a long time, can decelerate the motions of intruding stars
by dissipative forces \citep{Ostriker1983} and enhance the growth of the BH \citep{Just2012,Kennedy2016, Panamarev2018}.
Primordial binaries, detailed stellar evolution and stellar wind and partial tidal disruptions can 
also modify our results.
Including these mechanisms in the N-body simulations would help understand more detailed evolution of the IMBHs in star clusters.

\section{Summary}
\label{sec:summary}
%
We study the stellar dynamics around central IMBHs within early star clusters 
hosted by DM halos.
The IMBHs grow by TDEs of intruding stars, to become as massive as
$700\,\msun$ to $2500\,\msun$ at the end of the simulations. 
The diversity can be attributed, partly, to the variety of physical properties
of the parent clusters. 
Specifically, we find the final IMBH mass $M_{\rm IMBH,f}$ is approximately
linearly proportional
to the cluster mass $M_{\rm cl}$, yielding a relation $M_{\rm IMBH,f}\sim 0.01-0.02M_{\rm cl}$.
The TDE rates by the IMBHs are $\dot{N}_{\rm TDE}\sim0.3-1.3\,\Myr^{-1}$.
We show the rates follow the relations $\dot{N}_{\rm TDE}\sim0.3\,\Myr^{-1}\,(M_{\rm IMBH}/1000\,\msun)^2$.
The disrupted stars have masses in the range from $\sim3\,\msun$ to $\gtrsim100\,\msun$.
These are close to the minimum and maximum mass of the IMF $m_{\rm min}=3\,\msun$
and $m_{\rm max}=100\,\msun$, respectively.
The high mass stars are disrupted typically after migration toward the cluster center
through mass segregation.

The DM halo affects the evolution of $\mclb$ in a complicated manner (\figref{fig:bmass}).
Comparing our model A simulations with and without DM,
we find that $\mclb$ decreases when the DM density increases
and effectively ``heats'' the star cluster. 
By increasing the stellar velocity dispersion,
the DM halo strips stars from the outer part, but
stars in the dense inner region are not significantly affected by the DM.
We also find that the DM can prevent ejections of stars from the parent halo.
TDE rates are also affected by the DM motion.
Massive stars are stripped from the outer part of the cluster,
and mass segregation, which effectively promotes TDEs,
is supressed.

Though the IMBHs formed in the star clusters would not grow solely by the TDEs, 
they are still promising candidates for seeding SMBHs at high redshift $z\gtrsim 6$. 
Studying further IMBH evolution with halo mergers in a
cosmological context will help understanding the fates of the early IMBHs.

\section*{Acknowledgements}
We are grateful to Kazumi Kashiyama and Kojiro Kawana for discussions on our estimate of TDE rates.
The computations in this work are carried out on Cray XC30 and XC50 at CfCA, National Astronomical Observatory of Japan.
This study is financially supported by Grant-in-Aid for JSPS Overseas Research Fellowships (YS)
and JSPS KAKENHI Grant number 2680108, 17H6360 (MF).

\small{\bibliography{ms}}

\end{document}